\documentclass[1p,times]{elsarticle}

\usepackage{amsmath} 
\usepackage{amssymb}
\usepackage{graphicx}
\usepackage{subfigure}
\usepackage{multirow}
\usepackage{url}

\journal{An international Journal}

\usepackage{paralist}
\usepackage{bbding}
\usepackage{soul, color}
\usepackage{xspace}

 \usepackage{algorithmic}
  \usepackage{algorithm}

\usepackage{verbatim}
\usepackage{multirow}
\usepackage{multicol}

\usepackage{framed}

\usepackage{booktabs}

\definecolor{mk}{RGB}{50, 50, 250}

\def\dlf{\ifmmode{\cal DLF}\else${\cal DLF}$\fi}   
\def\dld{\ifmmode{\cal DLFD}\else${\cal DLFD}$\fi} 
\def\dlcfd{\ifmmode{\cal CFD}\else${\cal CFD}$\fi} 

\def\dlfp{\ifmmode{\mathit{partial-}\cal DLF}\else$\mathit{partial-}{\cal DLF}$\fi} 
\def\dldp{\ifmmode{\mathit{partial-}\cal DLFD}\else$\mathit{partial-}{\cal DLFD}$\fi} 
\def\dlcfdp{\ifmmode{\mathit{partial-}\cal CFD}\else$\mathit{partial-}{\cal CFD}$\fi} 

\RequirePackage{xspace}

\newcommand{\set}[1]{\{ #1 \}}          

\newcommand{\KM}
        {\ensuremath{\mathbf{K}}\xspace}
\newcommand{\KT}
        {\ensuremath{\mathbf{KT}}\xspace}
\newcommand{\Kfour}
        {\ensuremath{\mathbf{K4}}\xspace}
\newcommand{\Sfour}
        {\ensuremath{\mathbf{S4}}\xspace}
\newcommand{\Km}
        {\ensuremath{\KM_{\mathbf{m}}}\xspace}
\newcommand{\KTm}
        {\ensuremath{\KT_{\mathbf{m}}}\xspace}
\newcommand{\Kfourm}
        {\ensuremath{\mathbf{K4}_{\mathbf{m}}}\xspace}
\newcommand{\Sfourm}
        {\ensuremath{\mathbf{S4}_{\mathbf{m}}}\xspace}

\newcommand{\I}{\mathcal{I}}           

\newcommand{\K}{\mathcal{K}}                         

\newcommand{\IsSubs}{\sqsubseteq}          

\newtheorem{definition}{\textbf{Definition}}

\DeclareMathAlphabet\mathbfcal{OMS}{cmsy}{b}{n}

\newcommand{\dlrus}{\ensuremath{\mathcal{DLR}_{\mathcal{US}}}\xspace}
\newcommand{\DLRUS}{\ensuremath{\mathcal{DLR}_{\mathcal{US}}}\xspace}

\newcommand{\dlrifd}{\mathcal{DLR}_{\mbox{\emph{{\fontfamily{phv}\selectfont{\footnotesize ifd}}}}}}

\newcommand{\ERVTpp}{\ensuremath{EER{_{VT}^{++}}}\xspace}
\newcommand{\ERVT}{\ensuremath{ER{_{VT}}}\xspace}

\newcommand{\ourERT}{{\sc {\sc Trend}}\xspace}

\def\U   {\mathbin{\mathcal{U}}}                
\def\S   {\mathbin{\mathcal{S}}}                

\def\TS {\ensuremath{\mathcal{T}}\xspace}
\def\B  {\ensuremath{\mathcal{B}}\xspace}

\def\per{\textbf{.}\xspace}

\def\IsSubs{\sqsubseteq}

\newcommand{\ALlang}[2]{\ensuremath{\mathcal{#1}#2}\xspace}

\newcommand{\LS}{\ALlang{L}{}}
\newcommand{\ES}{\ALlang{E}{}}
\newcommand{\CS}{\ALlang{C}{}}
\newcommand{\RS}{\ALlang{R}{}}
\newcommand{\AS}{\ALlang{A}{}}
\newcommand{\US}{\ALlang{U}{}}
\newcommand{\DS}{\ALlang{D}{}}
\newcommand{\IS}{\ALlang{I}{}}

\newcommand{\FS}{\ALlang{F}{}}

\newcommand{\CST}{\ALlang{C}{^{T}}}
\newcommand{\RST}{\ALlang{R}{^{T}}}
\newcommand{\AST}{\ALlang{A}{^{T}}}

\newcommand{\CSS}{\ALlang{C}{^{S}}}
\newcommand{\RSS}{\ALlang{R}{^{S}}}
\newcommand{\ASS}{\ALlang{A}{^{S}}}

\newcommand{\CSM}{\ALlang{C}{^{M}}}

\newcommand{\att}{\mbox{\sc att}\ALlang{}{}}
\newcommand{\rel}{\mbox{\sc rel}\ALlang{}{}}
\newcommand{\crd}{\mbox{\sc card}\ALlang{}{}}

\newcommand{\cmin}{\mbox{\sc cmin}\ALlang{}{}}
\newcommand{\cmax}{\mbox{\sc cmax}\ALlang{}{}}

\newcommand{\isa}{\ensuremath{\mathbin{\mbox{\sc isa}\ALlang{}{}}}}
\newcommand{\ident}{\mbox{\sc id}\ALlang{}{}}
\newcommand{\disj}{\ensuremath{\mathbin{\mbox{\sc disj}\ALlang{}{}}}}
\newcommand{\cover}{\ensuremath{\mathbin{\mbox{\sc cover}\ALlang{}{}}}}

\newcommand{\player}{\mbox{\sc player}\ALlang{}{}}
\newcommand{\role}{\mbox{\sc role}\ALlang{}{}}

\newcommand{\as}{\ensuremath{\mathbin{\mbox{\sc s}\ALlang{}{}}}}
\newcommand{\at}{\ensuremath{\mathbin{\mbox{\sc t}\ALlang{}{}}}}

\newcommand{\chg}{\ensuremath{\mathbin{\mbox{\em chg}\ALlang{}{}}}}
\newcommand{\ext}{\ensuremath{\mathbin{\mbox{\em ext}\ALlang{}{}}}}
\newcommand{\mchg}{\ensuremath{\mathbin{\mbox{\em mchg}\ALlang{}{}}}}
\newcommand{\mext}{\ensuremath{\mathbin{\mbox{\em mext}\ALlang{}{}}}}
\newcommand{\qchg}{\ensuremath{\mathbin{\mbox{\em qchg}\ALlang{}{}}}}
\newcommand{\qext}{\ensuremath{\mathbin{\mbox{\em qext}\ALlang{}{}}}}
\newcommand{\mqchg}{\ensuremath{\mathbin{\mbox{\em mqchg}\ALlang{}{}}}}
\newcommand{\mqext}{\ensuremath{\mathbin{\mbox{\em mqext}\ALlang{}{}}}}
\newcommand{\CHG}{\ensuremath{\mathbin{\mbox{\sc chg}\ALlang{}{}}}}
\newcommand{\EXT}{\ensuremath{\mathbin{\mbox{\sc ext}\ALlang{}{}}}}
\newcommand{\MCHG}{\ensuremath{\mathbin{\mbox{\sc Mchg}\ALlang{}{}}}}
\newcommand{\MEXT}{\ensuremath{\mathbin{\mbox{\sc Mext}\ALlang{}{}}}}
\newcommand{\QCHG}{\ensuremath{\mathbin{\mbox{\sc Qchg}\ALlang{}{}}}}
\newcommand{\QEXT}{\ensuremath{\mathbin{\mbox{\sc Qext}\ALlang{}{}}}}
\newcommand{\MQCHG}{\ensuremath{\mathbin{\mbox{\sc MQchg}\ALlang{}{}}}}
\newcommand{\MQEXT}{\ensuremath{\mathbin{\mbox{\sc MQext}\ALlang{}{}}}}
\newcommand{\PCHG}{\ensuremath{\mathbin{\mbox{\sc Pchg}\ALlang{}{}}}}
\newcommand{\PEXT}{\ensuremath{\mathbin{\mbox{\sc Pext}\ALlang{}{}}}}
\newcommand{\PCHGr}{\ensuremath{\mathbin{\mbox{\sc PchgR}\ALlang{}{}}}}
\newcommand{\PEXTr}{\ensuremath{\mathbin{\mbox{\sc PextR}\ALlang{}{}}}}
\newcommand{\achg}{\ensuremath{\mathbin{\mbox{\sc chgA}\ALlang{}{}}}}
\newcommand{\Qachg}{\ensuremath{\mathbin{\mbox{\sc QchgA}\ALlang{}{}}}}
\newcommand{\frz}{\ensuremath{\mathbin{\mbox{\sc Frz}\ALlang{}{}}}}
\newcommand{\chgr}{\ensuremath{\mathbin{\mbox{\em chgR}\ALlang{}{}}}}
\newcommand{\extr}{\ensuremath{\mathbin{\mbox{\em extR}\ALlang{}{}}}}
\newcommand{\mchgr}{\ensuremath{\mathbin{\mbox{\em mchgR}\ALlang{}{}}}}
\newcommand{\mextr}{\ensuremath{\mathbin{\mbox{\em mextR}\ALlang{}{}}}}
\newcommand{\qchgr}{\ensuremath{\mathbin{\mbox{\em qchgR}\ALlang{}{}}}}
\newcommand{\qextr}{\ensuremath{\mathbin{\mbox{\em qextR}\ALlang{}{}}}}
\newcommand{\mqchgr}{\ensuremath{\mathbin{\mbox{\em mqchgR}\ALlang{}{}}}}
\newcommand{\mqextr}{\ensuremath{\mathbin{\mbox{\em mqextR}\ALlang{}{}}}}
\newcommand{\CHGr}{\ensuremath{\mathbin{\mbox{\sc chgR}\ALlang{}{}}}}
\newcommand{\EXTr}{\ensuremath{\mathbin{\mbox{\sc extR}\ALlang{}{}}}}
\newcommand{\MCHGr}{\ensuremath{\mathbin{\mbox{\sc MchgR}\ALlang{}{}}}}
\newcommand{\MEXTr}{\ensuremath{\mathbin{\mbox{\sc MextR}\ALlang{}{}}}}
\newcommand{\QCHGr}{\ensuremath{\mathbin{\mbox{\sc QchgR}\ALlang{}{}}}}
\newcommand{\QEXTr}{\ensuremath{\mathbin{\mbox{\sc QextR}\ALlang{}{}}}}
\newcommand{\MQCHGr}{\ensuremath{\mathbin{\mbox{\sc MQchgR}\ALlang{}{}}}}
\newcommand{\MQEXTr}{\ensuremath{\mathbin{\mbox{\sc MQextR}\ALlang{}{}}}}

\newcommand{\Intt}[1]{#1^{\I(t)}}
\newcommand{\Intv}[1]{#1^{\I(v)}}
\newcommand{\Intw}[1]{#1^{\I(w)}}

\newcommand{\Bint}[1]{#1^{\cal I}\/}
\newcommand{\Bintt}[1]{#1^{{\cal I}(t)}\/}

\newcommand{\Bintx}[2]{#1^{{\cal I}(#2)}\/}
\newcommand{\act}{\Bint{\Delta_{D}}}
\newcommand{\acti}{\Bint{\Delta_{D_i}}}
\newcommand{\acta}{\Bint{\Delta}}
\newcommand{\acto}{\Bint{\Delta_{O}}}

\newcommand{\TSS}{\ensuremath{\mathcal{T}_{\!p}}\xspace}

\newcommand{\EXISTR}[3]{\exists^{{\scriptscriptstyle\lessgtr} #1} [#2]#3}

\def\nexttime{\mbox{\ensuremath\oplus\,}}
\def\prevtime{\mbox{\ensuremath\ominus\,}}
\def\alltimep{\Box^+}
\def\alltimem{\Box^-}
\def\alltimes{\Box^*}
\def\sometimep{\Diamond^+}
\def\sometimem{\Diamond^-}
\def\sometimes{\Diamond^*}

\newcommand{\selects}[2]{#1 : #2}

\newcommand{\auf}{\left\langle}
\newcommand{\zu}{\right\rangle}

\newcommand{\schema}{\Sigma\ALlang{}{}}
\usepackage{tabularx}

\begin{document}

\begin{frontmatter}

\title{The temporal conceptual data modelling language {\sc Trend}}

\author[aff1]{Sonia Berman}
\author[aff1]{C. Maria Keet\corref{authorinfo}}
\ead{mkeet@cs.uct.ac.za}
\author[aff1]{Tamindran Shunmugam}
\cortext[authorinfo]{Corresponding author.  ORCID: \url{https://orcid.org/0000-0002-8281-0853}}

\address[aff1]{Department of Computer Science, University of Cape Town, South Africa}

\begin{abstract}
Temporal conceptual data modelling, as an extension to regular conceptual data modelling languages such as EER and  UML class diagrams, has received intermittent attention across the decades. It is receiving renewed interest in the context of, among others, business process modelling that needs robust expressive data models to complement them.
None of the proposed temporal conceptual data modelling languages have been tested on understandability and usability by modellers, however, nor is it clear which temporal constraints would be used by modellers or whether the ones included are the relevant temporal constraints.
We therefore sought to investigate temporal representations in temporal conceptual data modelling languages, design a, to date, most expressive language, \ourERT, through small-scale qualitative experiments, and finalise the graphical notation and modelling and understanding in large scale experiments. 
This involved a series of 11 experiments with over a thousand participants in total, having created 246 temporal conceptual data models. 
Key outcomes are that choice of label for transition constraints had limited impact, as did extending explanations of the modelling language, but expressing what needs to be modelled in controlled natural language did improve model quality.
The experiments also indicate that more training may be needed, 
in particular guidance for domain experts, 
to achieve adoption of temporal conceptual data modelling by the community.
\end{abstract}

\begin{keyword}
Entity-Relationship Diagrams; Temporal Conceptual Data Modeling; User Studies
\end{keyword}

\end{frontmatter}

\section{Introduction}

Temporal constraints in conceptual data models are important for many application domains, ranging from temporal conceptual data modelling, to business process modelling, data analytics, and designing append-only databases for blockchains, among others. Such a variety of application scenarios would ideally be served by a very expressive temporal conceptual data modelling language (TCDML) that would then be used throughout the various application scenarios for 
once-off 
effort of learning as well as easy reuse. Most extant TCDMLs, however, either lack transition constraints for object migration, temporal attributes, or relation migration (among others, \cite{Artale07a,AK08dl,Gregersen99,Khatri14,Ongoma15}), and none of them has been evaluated by users on preferences for diagram notation with, importantly, temporal constraints, nor for understanding of 
a
temporal model or ability to create a temporal conceptual data model. Here we 
include
modellers in industry 
as well as undergraduate and postgraduate students in academia who have been taught conceptual  modelling. 
Thus, there is, as a minimum, a need for an evaluation on fitness for purpose.

In this paper, we aim to address this through the development of a TCDML that is grounded in the description logic language $\dlrus$ and for which we designed and evaluated a graphical notation through a series of experiments. 
The resulting TCDML we called \ourERT,  Temporal information Representation in ENtity-relationship Diagrams.
In so doing, 
 we sought to answer the following questions:
\begin{itemize}
	\item[RQ1:] What diagram notation is preferred for temporal elements and constraints?
    \item[RQ2:] How well is the resulting temporal conceptual model 
    understood by modellers?
    \item[RQ3:] How well are modellers able to design such models?
\end{itemize}
To answer these questions, we conducted three sets of experiments. The first was a series of progressive experiments to determine the `best', or most preferred, notation of temporal information among the plethora of notations, using a qualitative reflective learning approach with participants mainly from industry and a moderately-sized evaluation with postgraduate students. 
The second set of experiments focussed on model understanding and feedback on a few minor variants in notation for the transition constraint, among postgraduate students and a large cohort of undergraduate students.  
The third set of experiments zoomed in on the ability to design temporal conceptual models with 
a temporal conceptual data modelling language we designed, 
through varying the amount of explanation and the task to carry out, in three large-scale experiments with 200-450 participating students that are 
comparative
novices in modelling, for increased statistical validity of the conclusions. 

The qualitative evaluations led to 
 a `tentative' notation for our TCDML, \ourERT, and 
also indicated a need for some training on modelling temporal information. The  quantitative experiments finalised and solidified the \ourERT notation. The experiments revealed that the labels for the transition constraints have little impact on correctness, but a graphical notation is clearly preferred over controlled natural language text. Temporal entity types, attributes, and dynamic extension are best used. More extensive introductory notes did not assist substantially in modelling and understanding overall, although a marked increase 
in 
modelling in temporal attributes and relationships was observed. Better demarcation of the domain, in the form of a set of controlled natural language statements, did improve modelling quality of the temporal constraints.

This paper extends our previously published work \cite{Keet17creol,KB17} in a number of ways. While \ourERT was tested with a small number 
of 
participants in \cite{KB17}, here we summarise also the experiments leading up to that version and conduct large scale evaluations that also finalised the graphical notation. In addition, the preliminary \ourERT was only sketched in \cite{KB17}, whereas here we present the full language---graphically, textually, and with its formal logic-based underpinning for precision and reference. 
We also provide an overview of the related work more comprehensively.

The remainder of the paper is structured as follows. Section~\ref{sec:relwork} summarises related work. Section~\ref{sec:designExp} describes the first set of experiments and results to determine the preferred representation in extended ER diagrams. Section~\ref{sec:TrendSpec} and Appendix~\ref{app:logic:ourERT} introduce \ourERT and its formal foundation. Section~\ref{sec:eval} reports on the experiments on model understanding and ability to design.
Section~\ref{sec:disc} discusses the 
work
and Section~\ref{sec:concl} concludes.

\section{Related work}
\label{sec:relwork}

The related work is divided into three sections. We begin with laying out a few theoretical principles relevant to the context. The two successive sections describe the extant temporal conceptual modelling languages that have a graphical and/or textual representation, and some further notions on how to represent temporal information based on related visualisation work. The latter has been added to broaden the view, because the former did not demonstrate much creativity in representation options. 

\subsection{Theoretical principles on graphical language design}
\label{sec:theo}

We first lay out four theoretical principles that will be adopted and extrapolated into the experiments so as to get a better understanding of possible preferred temporal information representation.

The first principle is that of of usability. This principle, as per definition in the ISO 9241-11 standard, is ``the extent to which a product can be used by specified users to achieve specified goals with effectiveness, efficiency and satisfaction'' \cite{Baxter05} in a specified context of use. This principle will be experimented with in terms of the ability to use the proposed notation when communicating the conceptual model during the Software Development Lifecycle (SDLC).

The second principle is that of notation, or ``visual notation'', as defined by Moody \cite{Moody09}. He describes that the visual notation can also be represented as visual, graphical, or diagramming notation consisting of a set of: 1) graphical symbols (the visual vocabulary), 2) compositional rules (the visual grammar), and 3) definitions of the meaning of each symbol (the visual semantics).
This research will highlight visual conceptual models and visual semantic components that are used to represent the temporal aspects of conceptual models and their perceived meanings they represent. 

The third principle is that of communication theory \cite{Shannon48}. Communication is a key concept for the interpretation of conceptual models to aid understandability. Most humans seemingly understand concepts better when they see how it works. 
The aim in this principle is to show that preferred visual metaphors help with, principally: communicating complex ideas effectively, that humans retain information longer when they understand what they are looking at, communicating to a wide range of people with different backgrounds, condensing information onto a limited canvas, and attract and hold the attention of observers as a learning tool.

Lastly, we take note of principles of the SDLC. That is, accepting that  visual conceptual models play an important role viz. in the communication, planning, and data analysis/modelling phases of software development\footnote{This is not to say that conceptual models are not used elsewhere, notably during runtime for graphical querying \cite{Catarci94,CKNRS10,Soylu17}, but that scenario is out of the current scope.}. 

\subsection{Visual Representations in temporal conceptual data models}

Many different TCDMLs have been introduced for the visualisation of temporal information for database development. In this short review, we will include also `older' TCDMLs, because they typically have more graphical adornments for temporal information than the more recent ones. It is of note that the vast majority of proposals have their origins in (Extended) Entity Relationship Diagram (ERD) rather than UML. Because of this and that the two approaches for extension proposed for UML Class Diagrams are not transferrable to ERD\footnote{To the best of our knowledge, there are stereotypes \cite{Cabot03} and OCL, such as in \cite{Ziemann03}, but no ERD variant has either of them.}, we focus on ERD-based modelling languages. They are summarised in Table~\ref{tab:modelVis} at the end of this section. The temporal aspects included may be valid time (true in the real world), transaction time (when the fact was stored in the database), or both, and more or less constraints on temporal classes, relationships, and attributes.   

For the older models proposed before 1999, we largely base the overview on Gregersen and Jensen's review \cite{Gregersen99}, because not all of the original papers were available. TERM \cite{Klopprogge81} was the first temporally extended ER with valid and transaction time and multiple granularities 
(as described in \cite{Gregersen99}). 
TERM is entirely text-based, as is TEER \cite{Elmasri90}, and has no adornments at all in the diagrams, unlike the 15 others. Gregersen and Jensen's `TempEER' (unnamed by the authors) is TEER extended to the relational model according to the proposers \cite{Lai94}, and thus obviously has no relevant adornments. 

\begin{figure}[t]
\centering
\includegraphics[width=0.7\textwidth]{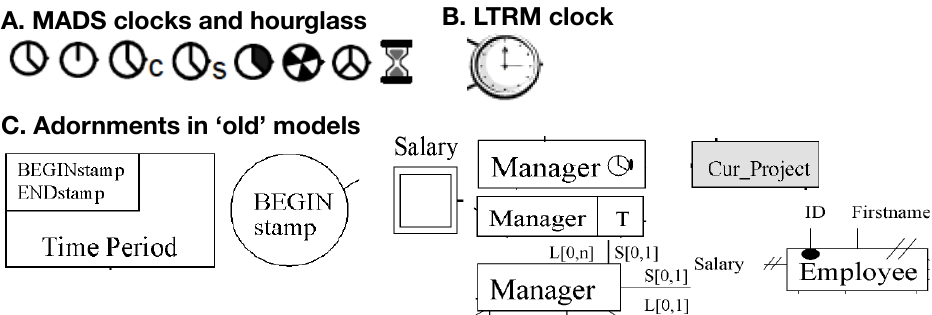}
\caption{A: all MADS icons for temporal elements; B: the LTRM clock; C: characteristic graphical adornments of the older models reviewed in \cite{Gregersen99}.}
\label{fig:oldGraphics}
\end{figure}

The other seven `old' TCDMLs all have one or more adornments to the ERD, of which only the most recent in that set had a clock icon, being TERC+ from 1997 \cite{Zimanyi97}. TERC+ is a predecessor to the spatiotemporal MADS \cite{Parent06} that was developed by the same group over the period of 1997-2006, which in its final version has seven different clock types and one hourglass to indicate different aspects of time, such as time, instant, interval, and timespan (see Fig.~\ref{fig:oldGraphics}-A). The only other TCDML with an icon, also a clock, is LTRM \cite{Mahmood10}, as depicted in Fig.~\ref{fig:oldGraphics}-B.

RAKE \cite{Ferg85} 
extends the ER model with timestamps to 
supports modelling 
the type DATE, which still persist in current DBMSs, and 
has further attributes such as a `BEGIN' and time periods as a class attached to a relationship. Other graphical extensions include signalling temporal elements by creating a double-lined element, as in MOTAR (Narasimhalu in \cite{Gregersen99}), shading temporal elements like in STEER \cite{Elmasri90}, and adding two parallel diagonal lines as in `TempRT' (Kraft in \cite{Gregersen99}). The characteristic graphics are included in Fig.~\ref{fig:oldGraphics}-C.

\begin{figure}[t]
\centering
\includegraphics[width=0.95\textwidth]{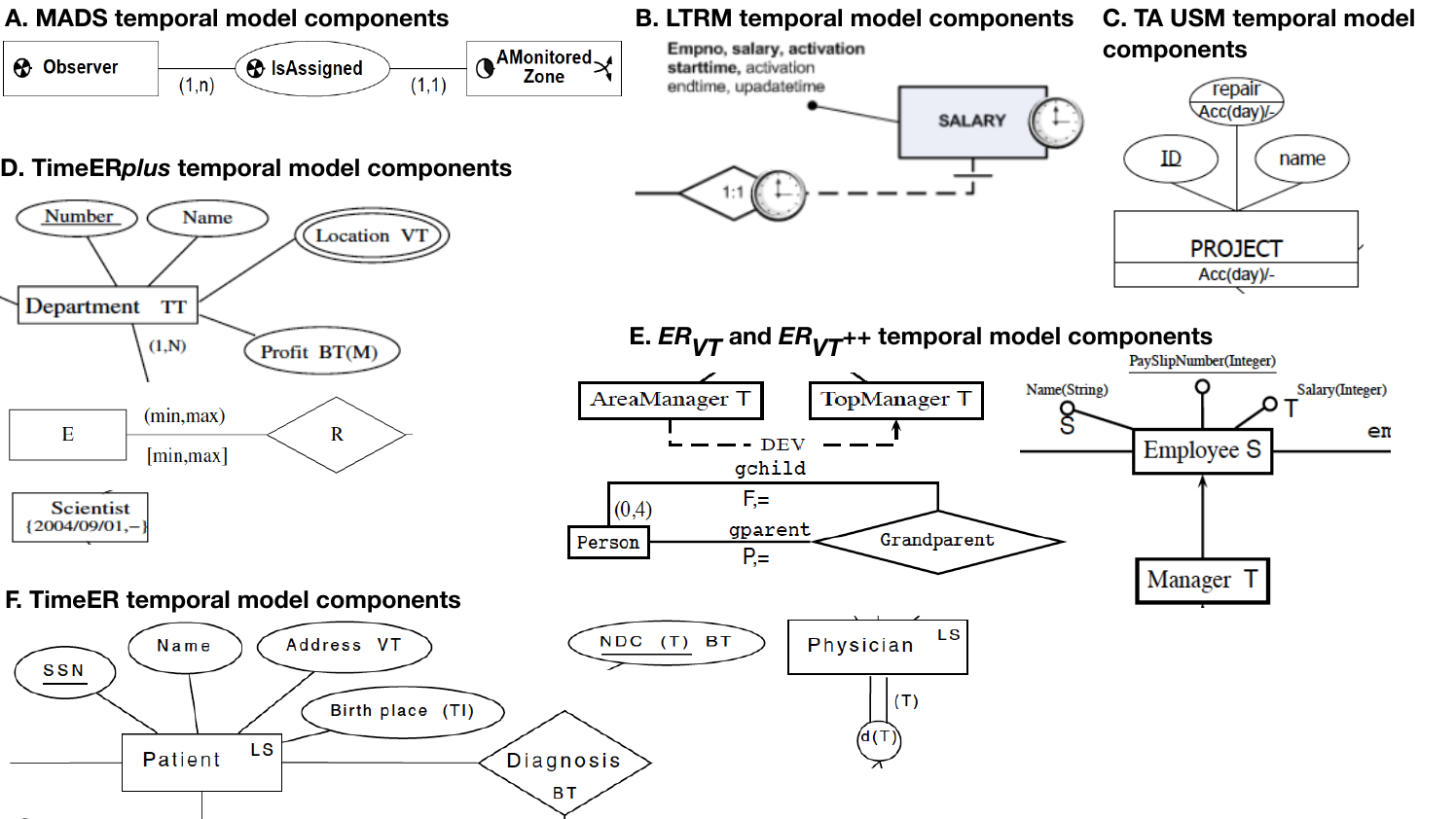}
\caption{A selection of model elements from six different TCDMLs to illustrate the sort of adornments used for temporal constraints in the `recent' proposals.}
\label{fig:modelEx}
\end{figure}

TER by Tauzovich, as described in \cite{Gregersen99}, differentiates itself by including lifetime and snapshot time varying cardinality constraints. The lifetime aspects are denoted by an ``L'' with {\em minL} and {\em maxL} and snapshot by ``S'' with {\em minS} and {\em maxS} and they are represented on the model using the following format: {\tt L [minL,maxL]} and {\tt S[minS,maxS]}. A similar adornment is used in TimeER{\em plus} \cite{Gregersen05}. This brings us to the more recent proposals which, as can be seen also from Table~\ref{tab:modelVis}, have predominantly textual adornments rather than graphical modifications or icons, with the exception of LTRM and MADS. A selection of the typical notations in context are included in Fig.~\ref{fig:modelEx}.

LTRM is based on mathematical expressions \cite{Mahmood10} where data is represented with past as well as present values and {\em activation\_start} and {\em activation\_end} on a temporal entity that is indicated with a clock symbol. That is, it has a combination of ways to communicate temporal information. 

Three of the eight `recent' (post-1999) models follow the same approach of adornments as ERT \cite{McBrien92}, which  adds a mini-rectangle to the rectangle of the entity type or relationship, where a {\tt T} is written. 
This ``T-stamping'' approach is also used  in {\sc TimeER} \cite{Combi08}, TimeER{\em plus} \cite{Gregersen05}, \ERVT \cite{Artale07a}, and \ERVTpp \cite{Ongoma15}. Each of these adds further `letter stamps', depending on what the TCDML supports, such as VT, TT, BT, and LS.

Further adornments largely depend on the type of temporal features the TCDML supports. For instance, TimeER{\em plus} \cite{Gregersen05} extends EER with both valid time and transaction time, lifespan, and time granularities, similar to {\sc TimeER} \cite{Combi08}. Khatri et al.'s telic/atelic extension to USM, here dubbed `TA USM' as it was nameless in \cite{Khatri14}, add the notion of accomplishments denoted with {\tt Acc()} in the model, among others, which are formally specified in an annotation grammar. These are all temporal features of the elements. \ERVT has the notion of transitions, were objects can extend or migrate from one entity type to another, called dynamic extension ({\sc Dex}) and dynamic evolution ({\sc Dev}), respectively. 
This is indicated with labelled arrows in the diagrams.  
\ERVTpp \cite{Ongoma15} extends \ERVT with attribute and relation migration \cite{KA10,KO15}, also indicated with similar arrows and corresponding abbreviations ({\sc ADex} and {\sc RDex} etc.). 

Table~\ref{tab:modelVis} lists the modelling languages examined with the way how it represents temporal features.  Observe that every model differs in terms of the temporal constructs using components of HCI such as different shapes, shading, and pictures, and only TERM and TEER are entirely text based. Except for ST USM, none of these theoretical models have been tested on users.

\begin{table}[h]
\small
	\centering
	\caption{Summary of TCDML's representation of temporal information. 
	}			
		\begin{tabular}{|p{2.5cm}|p{6.5cm}|p{2.1cm}|p{0.6cm}|}
\hline
{\bf Model (year)} & {\bf Brief description of the representation of the main temporal constructs} & {\bf Text, Icon, or Graphical} & \textbf{HCI eval.} \\
		\hline \hline
		\multicolumn{4}{|l|}{{\em Older TCDMLs (included in the survey of \cite{Gregersen99})}}\\ \hline
		 TERM (1983) & Text based & N/A & No \\ \hline
		 RAKE (1985) & Timestamps and periods in text, extra rectangles and circles & Graphic and text & No\\ \hline
		 MOTAR (1988) & Double-lined diamonds and squares & Graphic & No\\ \hline
		 TEER (1990) & Text-based, separate list & N/A & No \\ \hline
		 TER (1991) & Lifetime and Snapshot times added & Text & No\\ \hline		 
		 ERT (1992) & Extended rectangles with T `time box'  & Graphic and text & No\\ \hline
		 STEER (1993) & Shaded circles and ellipses, separate text & Graphic & No \\ \hline	
		 `TempEER' (1994) & None (TT and VT added in relational model) & N/A & No \\ \hline		 
		 `TempRT' (1996) & Diagonal lines & Graphic & No\\ \hline		
		 TERC+ (1997) & Clock added to an element & Icon & No\\ \hline	\hline			 
		\multicolumn{4}{|l|}{{\em More recent TCDMLs}}\\ \hline		 	 
		 ST USM (2004) & Formal annotation schema & Text & Yes\\ \hline	
		 TimeER{\em plus} (2005) & VT , TT, BT etc, dates, time granularity, lifespan & Text & No\\ \hline	
		 MADS (1997-2006) & Different clocks and one hourglass on the temporal feature & Icons & No\\ \hline	
		 \ERVT (2007) & T for Temporal, S snapshot, DEV/DEX-labeled arrows for transitions & Text & No\\ \hline
		 {\sc TimeER} (2008) & Text, with VT, LS etc. & Text & No\\ \hline
		 LTRM (2010) & Clock on the temporal element, start/end text & Icon and text & No\\ \hline
		 `TA USM' (2014) & Formal annotation schema & Text & No\\ \hline
		 \ERVTpp (2015) &  As for \ERVT, also for attributes and relationships & Text & No\\ \hline  \hline
		\multicolumn{4}{|l|}{{\em The TCDML proposed in this paper}}\\ \hline
		 \ourERT (2017) &  Clock, pin, DEV and DEX for transitions & Icons and text & Yes\\ \hline  		 	
		\end{tabular}
	\label{tab:modelVis}
\end{table}

\subsection{Techniques for Temporal Data Visual Representations}

Because of the limited creativity on representing termporal notions, we describe other techniques of temporal data using graphical representations namely, Pointwise Object Browsing, Concentric Circles, Agenda, Time Line Browser, Lifelines, Knave, and AsbruView. The reason for broadening the scope is twofold: the adornments described in the previous section assume a lot of in-depth knowledge of temporal information that may not be familiar to the non-expert temporal modeller, and those proposals may perhaps have been limited to stay within the ER notation of ideas and ideas of related extensions, such as the clocks and the `letter stamping'. 
Several notations have been discussed in \cite{Shunmugam16}, such as the visualisation as  Lifelines \cite{Plaisant96}, Time Line Browser \cite{Cousins91} and ``time slider'' \cite{Kapler05}, Knave \cite{Shahar00}, Concentric Circles \cite{Carlis98,Daassi00}, and Agenda \cite{Daassi00}, which are summarised in Table~\ref{tab:historyOverview}, and we highlight three here for their distinct approaches.

\begin{figure}[h]
\centering
\includegraphics[width = 0.9\textwidth]{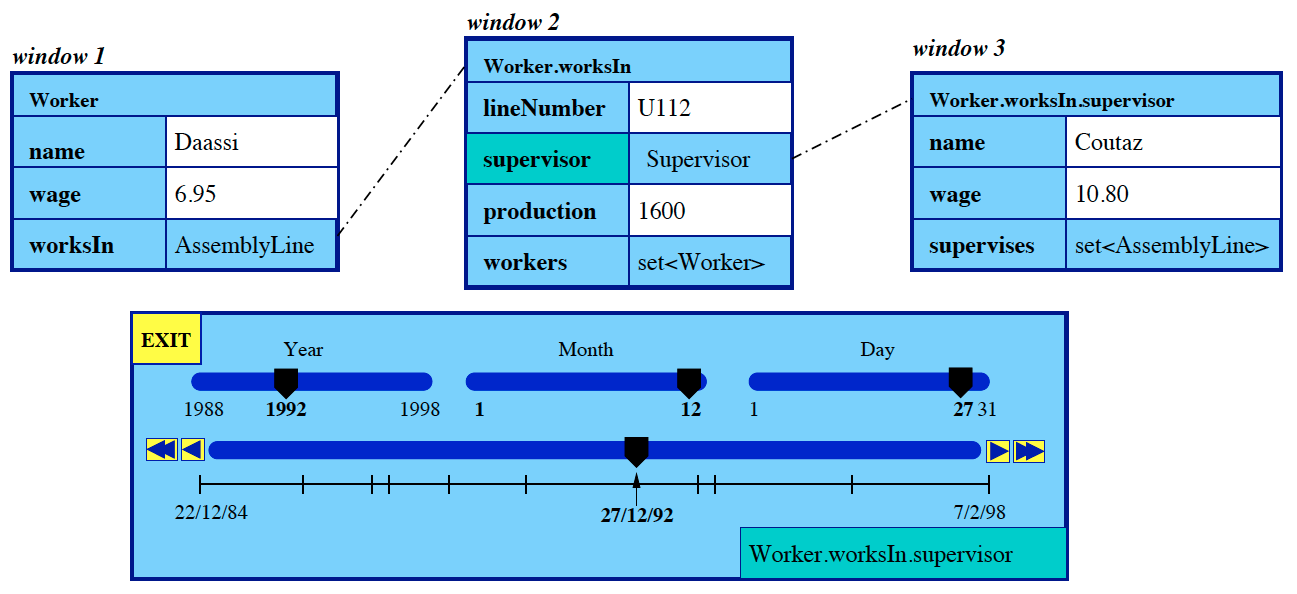}
\caption{Pointwise Object Browsing: representing temporal data in stages through a `drill down' traversing feature. (Source: \cite{Dumas01})}
\label{fig:POB}
\end{figure}

The Pointwise Object Browsing representation \cite{Dumas01} is used to display snapshots of temporal data at a given instant in a drill-down fashion. There are two ways for users to interact with the temporal data: by traversing through traditional object browsers or by altering the reference instant; a diagram is included in Fig.~\ref{fig:POB}. The features of this technique are that it is not sequential, it supports value alteration, and provides interaction devices for navigating to the previous or next instant where a modification arises in a given conception track expression. This may perhaps be useful for instance-motivated temporal modelling, like is possible for (atemporal) Object-Role Modeling \cite{Halpin08} and the FCO-IM method \cite{Bakema05} (cf. the natural language text-oriented temporal data \cite{Gianni14} or constraints \cite{Halpin08orm} in ORM that emphasise verbalisation of the information in pseudo-natural language sentences).

\begin{figure}[h]
\centering
\includegraphics[width = 0.9\textwidth]{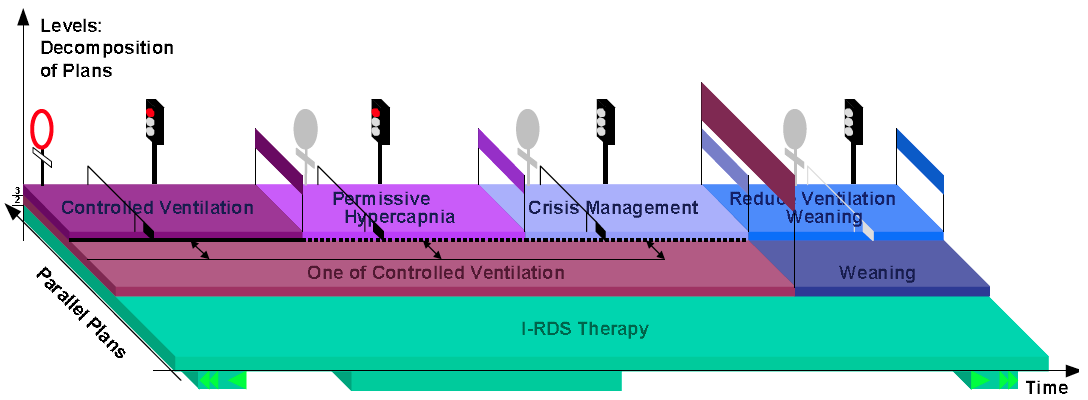}
\caption{An example of a different representation of temporal intervals and events with AsbruView. (Source: \cite{Kosara02})}
\label{fig:asbru}
\end{figure}

AsbruView \cite{Kosara02} deviates from any TCDML notation by representing `plans' in a 3-dimensional view, as to what is allowed to happen concurrently, what sequentially, and crucial change points. A sample diagram is included in Fig.~\ref{fig:asbru}.

Third, to be able to grasp the more complex notions of temporal information in temporal conceptual models, one needs to understand the simplified time period visualisation of intervals. Chittaro and Combi \cite{Chittaro01} looked into that with three different representations, using metaphors like springs and elastic bands. For instance, the end of strips must be connected and fixed by screws: an interval ends in this case, have fixed and set positions in relation with other intervals. The principal reason  motivating  this choice is that ``the fixed ends cannot move so the strip's ends are fixed.'' \cite{Chittaro01}. An example of the notation is included in the Appendix below (Fig.~\ref{fig:Exp2q}), for it was used in one of the questionnaires.

Table~\ref{tab:historyOverview} summarises the surveyed techniques for the representation of temporal data and history. Some techniques are purely visual whilst others combine text and visual indicators.

\begin{table}[h]
\small
	\centering
	\caption{Summary of Temporal Data History Representations. 
	}			
		\begin{tabular}{|p{2.6cm}|p{5.7cm}|p{3.8cm}|}
\hline
 & {\bf Temporal Constructs} & \textbf{Textual/Visual/Both} \\
		\hline \hline
		 Pointwise Object Browsing Technique & Uses Linear visualisation to show events at different instants of time & Visual linear with value alteration \\ \hline	 
		 Concentric Circles & Circles with common midpoint and different radii & Visual indicator size shows values over time \\ \hline	 		 
		 Agenda & Matrix based with line shows evolution of history over time period & Visual via hotspots \\ \hline	 
		Time Line Browser & Uses Linear visualisation to show events at different instants of time	 & Visual using snapshots \\ \hline	 
		 Lifelines & History based on different levels of granularity & Both text and visual indicators \\ \hline	 
		 AsbruView & Running track representation of history & Both text and visual indicators \\ \hline	 
		 Elastic Bands & Uses Linear visualisation to show events at different instants of time & Visual using bands to display events in time \\ \hline	 
		 Springs & Uses Linear visualisation to show events at different instants of time & Visual using springs to display events in time  \\ \hline	 
		 Paint Strips & Uses Linear visualisation to show events at different instants of time & Visual using paint strips to display events in Time
 \\ \hline	 		 		 		 		 		 		 
		\end{tabular}
	\label{tab:historyOverview}
\end{table}

\begin{table}[t]
\small
	\centering
	\caption{Summary of the experiments, in chronological order; n: number of participants, m: number of models; q: number of Multiple Choice Question} answers.\label{tab:Experiments}			
		\begin{tabular}{|p{0.35cm}|p{8.8cm}|p{2.85cm}|}
\hline
{\bf No.} & {\bf What} & {\bf Who}  \\ 	\hline \hline
E1 & Online survey (S1) on textual and graphical notations & compsci students, developers, db lecturers, business users (n=15) \\ \hline
E2 & Education, survey on interval notations, discuss \& educate & select group of business users (BU) (n=12) \\ \hline
E3 & Modelling change (survey), educate & BU (n=12) \\ \hline
E4 & Effects of training: S1 again + one question on preference, discussion & BU (n=12) \\ \hline
E5 & Model preferences survey  & BU (n=12) \\ \hline
E6 & Arrows \& labels; Modelling and understanding; questions on notation and on \ourERT vs text; no domain; in-class explanation & PG1 (n=m=15) \\ \hline
E7 & Modelling and understanding; questions on notation and on \ourERT vs text; Loans Company domain, restricted; in-class explanation & PG2 (n=m=14) \\ \hline
E8 & Dynamic constraint depiction (which labels), \ourERT vs text; Modelling and understanding; University domain, open; brief written memo on \ourERT & UG1 (n=177; m=43; q=170) \\ \hline
E9 & Dynamic constraint depiction ({\sc Dev}/{\sc Dex} vs {\sc Chg}/{\sc Ext}), \ourERT vs text; Modelling and understanding; University domain, open; extended written memo on \ourERT & UG2 (n=390; m=79; q=373) \\ \hline
E10 & Modelling and understanding (as translation to database schema); extended written memo on \ourERT and textbook basics on temporal databases; CNL statements in closed domain & PG3 (n=11)\\ \hline										
E11 & Dynamic constraint depiction ({\sc Dev}/{\sc Dex} vs {\sc Chg}/{\sc Ext}); Modelling and understanding; CNL statements in closed domain (Tourism); extended written memo on \ourERT  & UG3 (n=402; m=83) \\ \hline
\end{tabular}	
\end{table}

\section{Preferences for notations of temporal constraint in diagrams}
\label{sec:designExp}

To determine the preferred notation of temporal constraints that led up to {\sc Trend} and to evaluate modelling and understanding of them, a suite of experiments were conducted, which are enumerated and summarised in Table~\ref{tab:Experiments}. Those 11 experiments can be grouped into 3 stages: Experiment 1 (E1) to Experiment 6 (E6) that led up to a first assumed stable version of \ourERT; then a second stage with mainly  E7 and part of E8 that 
 had a dual function of finalising some notation and increasing
the scale to result in a possibly final \ourERT; and a third stage consisting mainly of (very) large scale experiments for increased statistical validity and robustness of conclusions to be drawn from them, consisting mainly of experiments E9 and E11 (with the small scale E10 as motivator for E11 to eliminate a lingering doubt emerging from the results of E9). From a research process perspective, experiments E1-E6 and part of E7 and E8 concern the {\em design} of the language, whereas the second set of experiments, being the other part of E7 and E8, E9, and E11, concern the {\em evaluation} of \ourERT. 
This section focuses on the 
experiments that influenced the design of \ourERT, whereas the large-scale evaluations are deferred to 
 Section~\ref{sec:eval}.

Based on the results of the first experiment, the methodology adopted was that of ``reflective teachings'' \cite{Steinbring98}, for most 
conceptual modellers
are not experts in temporal modelling. Reflective teachings amounts to: 
1) a set of tasks is presented to a group of users; 2) the results are critically analysed by the facilitator to ascertain the knowledge level of the population; 
3) If the 
level is too low, the facilitator conducts a knowledge transfer to the group to increase their knowledge level. 
This process may be repeated as often as needed. 
The main goals are to dispel all anxiety 
and ensure that the population knowledge level is increased to an extent whereby 
results 
can be used for effective analysis.

\subsection{E1-E5: reflective teachings experiments}

The set-up, results, and discussion about small-scale experiments E1-E5 are summarised in this section, with the details described in \cite{Shunmugam16}. The aim here is to provide an impression and insight in the process that led up to the draft notation of \ourERT, i.e., mainly serving to narrow down the plethora of possible graphical and textual adornments, rather than a detailed account for reproducibility. The successive experiments are outlined in their approach and key outcomes are discussed.

\subsubsection{Experiment 1: Textual and graphical notations} 

The aim of the initial experiment was to evaluate the representations of temporal information in conceptual models 
by means of an online questionnaire. 

This consisted of seventeen exercises and 
a section 
for feedback and comments. 
The URL was sent to an open audience 
of computer science students, developers, database lecturers, and business users, 
assumed 
to have
some knowledge of databases and ER diagrams. 
Some models used were
taken from existing work mentioned in Section~\ref{sec:relwork} and %
others 
had different colour schemes, 
visual icons, text and graphics. 
A selection of the diagrams is included in Fig.~\ref{fig:Exp1qs} in the Appendix. 
Subjects were asked their model preferences, and they were given one open question
requiring a description of what a small temporal conceptual model meant. 

\subsubsection{Experiment 2: Interval notations}

Results of Experiment 1 (E1) 
showed that 
few respondents 
understood the subject matter and temporal aspects in general. 
The aim of this, and the following, experiment was to close this knowledge gap. To gauge if the knowledge gap was indeed getting smaller, the 
audience had to be narrowed down to a smaller closed group 
subset of 12 business users with varied knowledge levels of information technology and database concepts.

Training 
was
provided before 
Experiment 2 (E2)
in the form of an online video explaining temporal aspects in Microsoft SQL Server 2016 \cite{MSvideo}. This was the first step to raise 
their 
knowledge level. 
Participants were then given 12 multiple choice questions to evaluate three temporal representations, 
viz.
springs, elastic bands and paint strips, as proposed in \cite{Chittaro01}. This
was followed by a classroom discussion, 
where
it was found that there was still 
confusion among the group about temporal data. Temporal concepts were further explained 
using the visual techniques of concentric circles, 
Agenda, 
time line browser, 
and AsbruView. 
They learnt about 
time intervals, transaction and valid time, history and why it is important, 
and differences between atemporal and temporal databases. When questioned about 
other difficulties
in E1, we found 
they also did not understand the proposed models. 
Hence, Experiment 3 was setup to introduce modelling to the group.

\subsubsection{Experiment 3: Modelling change} 

Three tasks were posed to the same participants as in E2, so as to gauge their understanding of modelling 
and to show 
that without a visual language for understanding conceptual models, 
interpretation of diagrams could be very diverse. 
The diagrams in Fig.~\ref{fig:Exp3q} (in the Appendix) were self-generated so as to be ambiguous, 
and the same 12 participants were asked to 
give their understanding of each.
There was no constraint on the time taken. 
Qualitative data was also recorded 
when subjects were asked to describe their interpretations verbally.

This showed 
that
some members 
did not grasp
concepts such as constraints, if-then, loops, dependencies, and flows.
The follow-up training 
explained
entities, attributes, migrations, both evolution and extension, constraints, inheritance and relationships, 
and
that a symbol on a diagram has different meanings depending on the context it is used in.

\subsubsection{Experiment 4: The effects of training} 

This was principally a repeat of E1, 
carried out on
the same, but now ``knowledgeable'', subjects. 
There was one minor change: 
an \ERVT, \ERVTpp, and a MADS model were added to the  
``Which conceptual model do you prefer?'' question.
In analysing the data, some discrepancies were observed, 
so a 
follow-up discussion was  
held
with the group 
to shed light on it.

Analysis of data from E4 
revealed
that the preferred temporal conceptual model 
was RAKE.
From interviews it was ascertained that this was chosen because of 
its simplified 
notation.
Some inconsistencies 
were noted however. There was a perceived preference for 
pictorial 
over 
textual representation, yet 
RAKE 
temporal indicators are text only. 
Secondly, the preferred 
picture for temporal representation 
was the clock, 
which is not used in RAKE.
On discussing this
with the group, it 
emerged that their preference  
was the ERD diagram and 
this outweighed the preference for the clock graphic. 
The
{\sc Dex}/{\sc Dev} textual representation was preferred by business users due to  
being easier to draw 
accurately 
on flipcharts and whiteboards, and 
easier to interpret.

\subsubsection{Experiment 5: Model preferences} 

In this experiment, the group was asked to rate 
temporal conceptual models
on a linear scale from 1 to 10.
All models were drawn with equivalent information.
Based on 
E4 discussions
and earlier stated preferences for
ERD models, for textual representation of migration constraints, and for less complexity, five questions were used in the survey.

\begin{figure}[h]
\centering
\includegraphics[width = 1.0\textwidth]{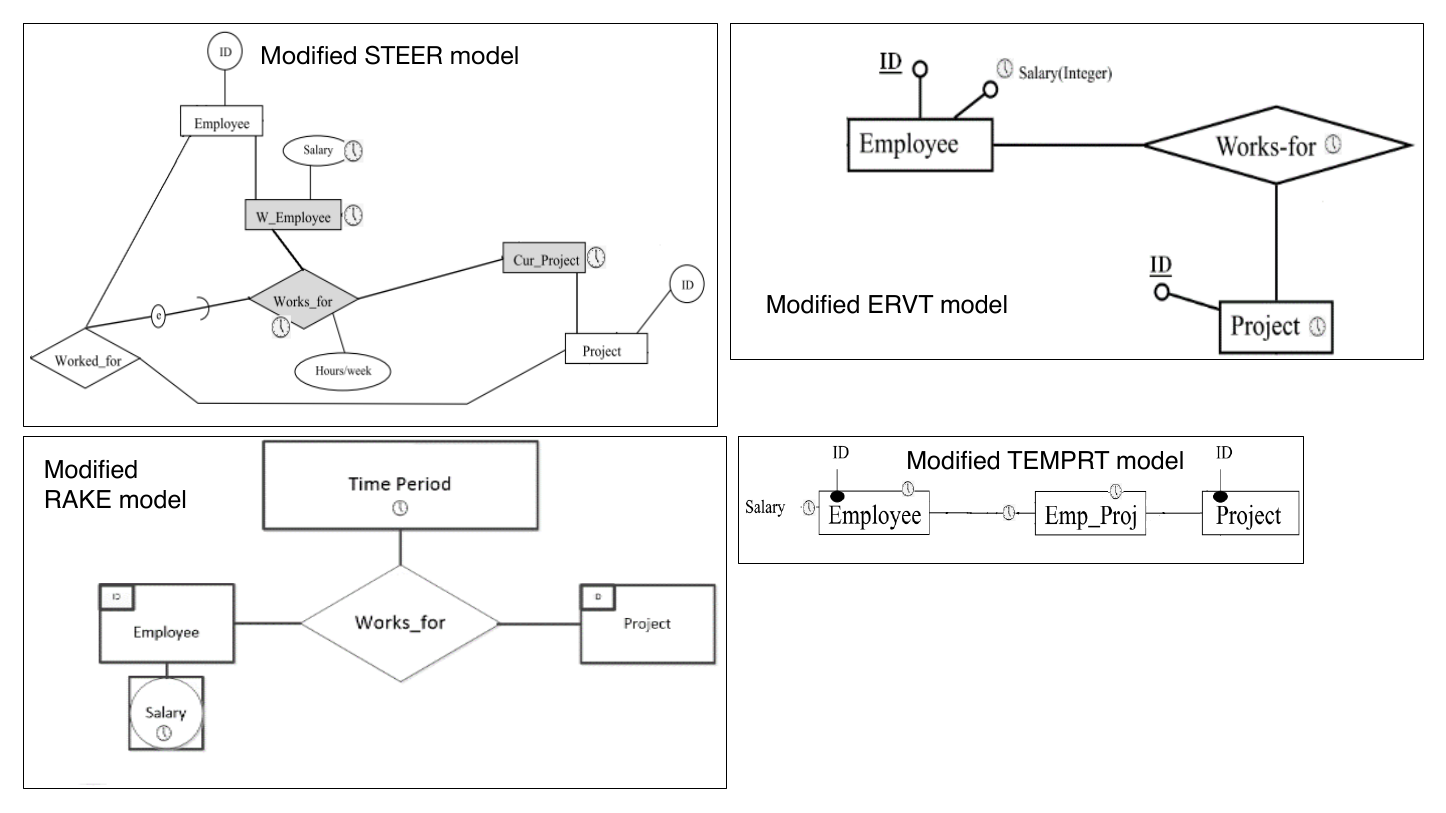}
\caption{Representation of the same temporal information in the six languages (only a selection is shown), updated with the clock preference, which participants had to rate.}
\label{fig:Exp5q2}
\end{figure}

\begin{figure}[h]
\centering
\includegraphics[width = 0.9\textwidth]{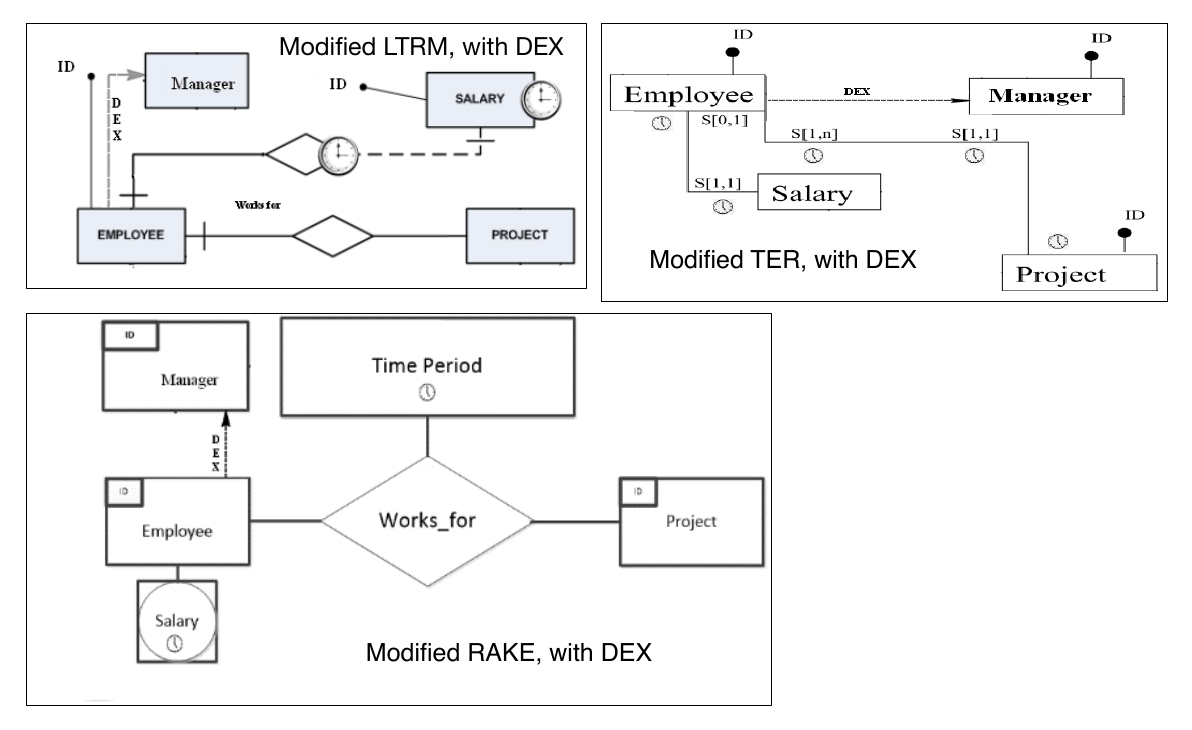}
\caption{Representation of the same temporal information in the six languages (only a selection is shown, which participants had to rate. Except for \ERVT that already has {\sc Dex} and {\sc Dev}, the other models have been extended with these constraints not present in the language.}
\label{fig:Exp5q34}
\end{figure}

Each model was derived from the ERD representation, namely: STEER, TER, LTRM, MOTAR, RAKE, \ERVT and TEMPPR models. 
The first survey question had them unmodified (Fig.~\ref{fig:Exp5q1} in the appendix), whereas the second 
had them modified using the clock icon (Fig.~\ref{fig:Exp5q2}).
All models (except for \ERVT) had 
a textual representation for constraint migration of {\sc Dex} in question 3,
and then also with the clock icon 
in question 4. A selection of the models is included in Fig.~\ref{fig:Exp5q34}. Comments about the survey were also elicited.

\subsection{Zooming in on dynamic constraint details with more subjects}

Finally, we wondered about the representation of dynamic (transition) constraints, not covered in the previous experiments. 
Questions were 
tagged onto  
two related experiments  
that are reported on in detail in Section~\ref{sec:eval}.
These 
involved three different, and larger, sets of university students, so as to canvas a broader group of potential users and potentially obtain results with more statistical validity and confidence.
Two kinds of transitions were described to participants: dynamic extension (abbreviated {\sc Dex}), being one where an entity remains an instance of its type but also becomes an instance of another type, and dynamic evolution (abbreviated {\sc Dev}), being one where an entity changes from being of one type to being of another type. 

\subsubsection{Experiment 6: Arrows and labels for dynamic constraints}

\begin{figure}[h]
\centering
\includegraphics[width = 0.7\textwidth]{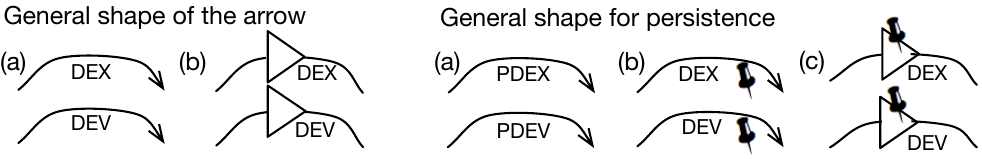}
\caption{Options for the basic arrow shape for the generic dynamic (transition) constraints and for persistence of those transitions.}
\label{fig:Exp6}
\end{figure}

Among a set of questions that will be described in detail in Sect.~\ref{sec:eval}, we evaluated preference for a labelled arrow to depict dynamic constraints, as opposed to one with a triangle as in UML's subsumption notation.
Fig.~\ref{fig:Exp6} shows the options that participants of E6 could choose from; {\sc Dev} and {\sc Dex} having been introduced as the labels representing the two types of dynamic transition. Additionally, we introduced dashed lines for optional transitions and solid lines for mandatory transitions, which was not asked about in the evaluation.

\subsubsection{Experiments 7 and 8 (part 1): transition labels}
In the other two experiments, E7 and E8, subjects were asked to propose alternative terms to  {\sc Dev} and {\sc Dex}, if they wished, as part of a set of questions (see Fig.~\ref{fig:MCQs}). 

\subsubsection{Experiment 9 (part 1): validating labels}
Results obtained from E7 and E8, which were held in the same year, led to another round of the experiment in a later cohort, 
comparing the labels {\sc Dev} and {\sc Dex} to labels {\sc Chg} and {\sc Ext}, respectively,
and, for dynamic constraints in the past, the ``$^-$'' superscript versus lower/uppercase, like {\sc Dex}$^-$ vs dex. The materials and methods are described in Sect.~\ref{sec:eval} (and only mentioned here for narrative flow), because they could be combined with the quantitive assessment since the class size had more than doubled in the meantime.

\subsubsection{Results}
Eleven of the 15 postgraduate students (PG1, in E6) answered the questions on
the arrow's shape in the related experiment \cite{KB17}. 
For the question on the arrow shape, 4 preferred the arrow and 7 preferred the triangle 
 as illustrated in Fig.~\ref{fig:Exp6}b. 
However, when put in the light of the question on how to denote persistent extension or evolution, the picture becomes more complicated. Four participants preferred as is, i.e., an arrow with {\sc PDex/PDev} (but they are not the same 4 as for the previous question), 3 preferred the pin next to the text,  2 preferred a pin in the triangle, and 2 ties for option b and c. Thus, now a majority does not want the triangle. Because of the ambiguous result, we keep the 
simpler notation, 
i.e., a simple arrow tip.

In the related experiment E8, involving 
179 participants (UG1), 100 of them provided 1892 alternative labels for {\sc Dex} and {\sc Dev}, 
with a substantial actual overlap in strings 
having the same stem. The most popular choices were:
{\sc Change} (13), {\sc Extend} (11), {\sc Add} (9), {\sc Becomes} (9), {\sc Also} (8),  
{\sc Transform} (8),
{\sc Mutate} (6) and {\sc Progress} (6).

Strings that had at least two proposers ($n=34$),  
were
independently categorised 
by two authors
as a possible alternative for {\sc Dev}, {\sc Dex}, or neither. Fourteen strings were categorised the same by both authors. Combining this with most-proposed,  
{\sc Extend} 
emerged as 
best alternative for {\sc Dex}, and {\sc Change} or {\sc Transform} 
for {\sc Dev}. 
These could be abbreviated as {\sc Ext}, and  
{\sc Chg} or {\sc Tran}. 

As we shall see in Sect.~\ref{sec:eval}, E9 showed that there is no overwhelming preference for one or the other set of labels for the transition constraints, nor does it have a large combined effect in modelling and understanding of \ourERT. We therefore proceeded to defining \ourERT (presented next, in Sect.~\ref{sec:TrendSpec}) and examine its usability, which will be reported on in Section~\ref{sec:eval}.

\section{Specification of \ourERT}
\label{sec:TrendSpec}

\begin{figure}[h]
\centering
\includegraphics[width = 0.95\textwidth]{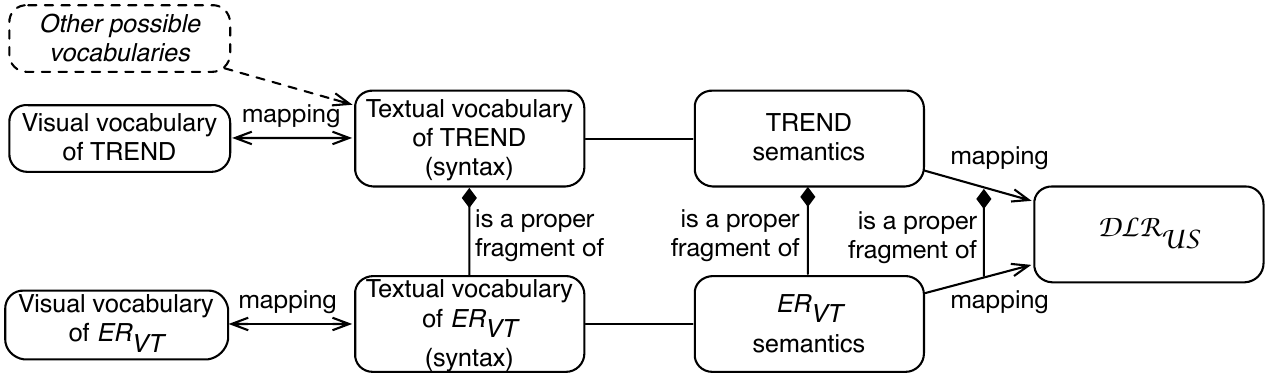}
\caption{Orchestration of approach to obtain the links between a swappable visual modelling language, the reusable syntax and semantics, and relation to \DLRUS, and their relation to the  \ERVT specification  in \cite{AK08is} (equally applicable for \ERVTpp). Observe the bi-directional mapping between text and graphics and the unidirectional mapping into \DLRUS.}
\label{fig:trendspec}
\end{figure}

We now describe the visual vocabulary and semantics of \ourERT.
We do this by relying on the formal foundation of \ERVT (and thus also the extension to \ERVTpp) and the mapping presented in \cite{AK08is} and \cite{Ongoma15}, and extend it for the addition cf. \ERVTpp, being the distinction between optional and mandatory transitions, which in the logic-based reconstruction is an additional existentially quantified relation. 
As in \cite{AK08is,Ongoma15}, the specification consists of several steps, whose orchestration is depicted in Fig.~\ref{fig:trendspec}. Observe that \ourERT still fully relies on \DLRUS \cite{Artale02} and does not go beyond it; we exploit its expressiveness better. We summarise the key features first, roughly in the way it was also introduced to the participants of the experiments. Then the textual syntax and the direct semantics are presented, whereas $\dlrus$ and the mapping of \ourERT into $\dlrus$ is relegated to the Appendix.

\subsection{Key features of \ourERT} \label{TrendSpec}

\ourERT focuses on temporalising classes, relationships and attributes, and specifying temporal transition constraints. The core transition constraints are dynamic extension 
 ({\sc Ext}) 
and dynamic evolution 
({\sc Chg}).  
 In an extension, the entity is also an instance of the other entity whereas with evolution, the entity ceases to be an instance of the source entity type. An example of extension is when an Employee becomes a Manager they remain an Employee, and of evolution is when a SeniorLecturer becomes an AssociateProfessor because then they are no longer a SeniorLecturer. 
A clock icon indicates a temporal element, and arrows labeled with 
 {\sc Chg} 
and 
 {\sc Ext}  
represent temporal transitions. A dashed arrow shaft denotes an optional transition and a solid shaft denotes a compulsory transition. Labels are 
chg  
and 
ext   
when the transition constraints refers to the past; e.g., a {\sf Graduate} must have been a {\sf Student} before. 
Both future and past transitions are needed when these transition constraints differ: e.g., {\sf Student} becoming a {\sf Graduate} in future is optional, but {\sf Graduate} having been a {\sf Student} in the past is compulsory. 
Any transition constraint can include a time quantity to indicate how much time must elapse before the transition can occur.  

The final notation is shown in Table~\ref{tab:elements} and in Table~\ref{tab:constraints}.

\def\imagetop#1{\vtop{\null\hbox{#1}}}

\begin{table}[!h]
\small
	\centering
	\caption{\ourERT~basic elements. The formal syntax is given in Definition~\ref{def:syntax}.}		\label{tab:elements}	
		\begin{tabular}{|>{\centering\arraybackslash}c|p{2.5cm}|p{7.5cm}|}
\hline
\multicolumn{1}{|c|}{{\bf Element Icon}} & \multicolumn{1}{p{2.5cm}|}{{\bf  Textual syntax}} & \multicolumn{1}{c|}{{\bf Comment}} \\
		\hline \hline
\imagetop{\includegraphics[width=0.13\textwidth]{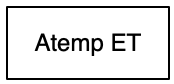}} & \raisebox{-0.4cm}{\parbox{2cm}{ \centering {\sc c} }}  & \raisebox{-0.4cm}{\parbox{7cm}{(regular) atemporal entity type (may also be written in full, with {\sc c}$^S$)}}  \\ \hline  	
\imagetop{\includegraphics[width=0.13\textwidth]{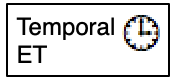}} & \raisebox{-0.4cm}{\parbox{2cm}{ \centering {\sc c}$^T$ }} & \raisebox{-0.4cm}{\parbox{6cm}{temporal entity type}}  \\ \hline  	
\imagetop{\includegraphics[width=0.13\textwidth]{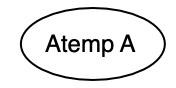}} & \raisebox{-0.4cm}{\parbox{2cm}{ \centering \att }} & \raisebox{-0.4cm}{\parbox{7.5cm}{atemporal attribute (may also be written in full, with \att$^S$) }}\\ \hline  	
\imagetop{\includegraphics[width=0.13\textwidth]{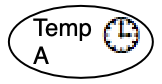}} &  \raisebox{-0.4cm}{\parbox{2cm}{ \centering \att$^T$ }} &  \raisebox{-0.4cm}{\parbox{6cm}{temporal attribute}} \\ \hline  	
\imagetop{\includegraphics[width=0.13\textwidth]{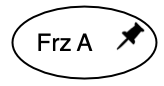}} & \raisebox{-0.4cm}{\parbox{2cm}{ \centering \frz }}  &  \raisebox{-0.4cm}{\parbox{6cm}{Frozen attribute}}  \\ \hline  	
\imagetop{\includegraphics[width=0.13\textwidth]{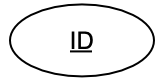}} & \raisebox{-0.4cm}{\parbox{2cm}{ \centering \ident }}  &  \raisebox{-0.4cm}{\parbox{6cm}{identifier attribute}} \\ \hline  	
\imagetop{\includegraphics[width=0.15\textwidth]{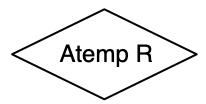}} & \raisebox{-0.4cm}{\parbox{2cm}{ \centering {\sc r} }} & \raisebox{-0.4cm}{\parbox{7.5cm}{atemporal relationship (may also be written in full, with {\sc r}$^S$)}} \\ \hline  	
\imagetop{\includegraphics[width=0.15\textwidth]{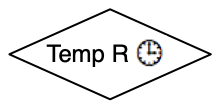}} & \raisebox{-0.4cm}{\parbox{2cm}{ \centering {\sc r}$^T$ }} & \raisebox{-0.4cm}{\parbox{6cm}{temporal relationship}}  \\ \hline  	
\imagetop{\includegraphics[width=0.02\textwidth]{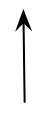}} & \raisebox{-0.4cm}{\parbox{2.5cm}{ \centering \isa$_C$, \isa$_R$, \isa$_U$}}  & \raisebox{-0.4cm}{\parbox{6cm}{Subsumption}}   \\ \hline  	
		\end{tabular}
\end{table}

\begin{table}[!h]
\small
	\centering
	\caption{\ourERT~constraints. The formal syntax is given in Definition~\ref{def:syntax}.}	\label{tab:constraints}	
		\begin{tabular}{|>{\centering\arraybackslash}c|p{2.8cm}|p{6.3cm}|}
\hline
\multicolumn{1}{|p{2.4cm}|}{{\bf Constraint Icon}} & \multicolumn{1}{p{3cm}|}{{\bf  Textual syntax}} & \multicolumn{1}{c|}{{\bf Comment}} \\
		\hline \hline
\imagetop{\includegraphics[width=0.05\textwidth]{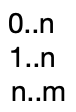}} & \raisebox{-0.4cm}{\parbox{3cm}{ \centering $\crd_R$ or $\crd_A$ }}  & \raisebox{-0.4cm}{\parbox{6cm}{cardinality on the relationship or the attribute (only a sample shown) }}  \\ \hline  	
\imagetop{\includegraphics[width=0.04\textwidth]{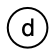}} & \raisebox{-0.4cm}{\parbox{2cm}{ \centering \disj$_C$ or \disj$_R$ }} & \raisebox{-0.4cm}{\parbox{6cm}{disjointness among subtypes or relationships }}  \\ \hline  	
\imagetop{\includegraphics[width=0.02\textwidth]{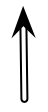}} & \raisebox{-0.4cm}{\parbox{2cm}{ \centering \cover }} & \raisebox{-0.4cm}{\parbox{6cm}{covering constraint for subtypes  }}\\ \hline  	
\imagetop{\includegraphics[width=0.2\textwidth]{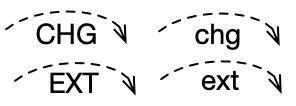}} & \raisebox{-0.5cm}{\parbox{2.8cm}{ \centering \CHG, \chg, \EXT, \ext, \CHGr, \chgr, \EXTr, \extr, \achg}} & \raisebox{-0.5cm}{\parbox{6.45cm}{optional change or extension, respectively, for in the future and in the past, respectively, for either entity types or relationships, and change for attributes}} \\ \hline  	
\imagetop{\includegraphics[width=0.2\textwidth]{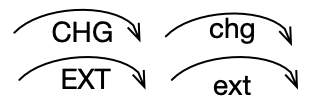}} & \raisebox{-0.4cm}{\parbox{3cm}{ \centering \MCHG, \mchg, \MEXT, \mext, \MCHGr, \mchgr, \MEXTr, \mextr }}  & \raisebox{-0.4cm}{\parbox{6cm}{mandatory change or extension, respectively, for in the future and in the past, respectively, for either entity types or relationships}}  \\ \hline  	
\imagetop{\includegraphics[width=0.2\textwidth]{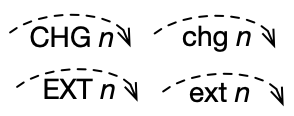}} & \raisebox{-0.4cm}{\parbox{3cm}{ \centering  \QCHG, \qchg, \QEXT, \qext, \QCHGr, \qchgr, \QEXTr, \qextr, \Qachg}}  & \raisebox{-0.4cm}{\parbox{6.45cm}{quantitative optional change or extension, respectively, for in the future and in the past, respectively, for either entity types or relationships, and change for attributes}} \\ \hline  	
\imagetop{\includegraphics[width=0.2\textwidth]{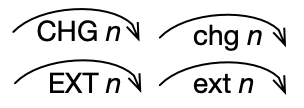}} & \raisebox{-0.4cm}{\parbox{3cm}{ \centering \MQCHG, \mqchg, \MQEXT, \mqext, \MQCHGr, \mqchgr, \MQEXTr, \mqextr}} & \raisebox{-0.4cm}{\parbox{6.45cm}{quantitative mandatory change or extension, respectively, for in the future and in the past, respectively, for either entity type or relationships}} \\ \hline  	
\imagetop{\includegraphics[width=0.1\textwidth]{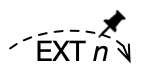}} & \raisebox{-0.4cm}{\parbox{3cm}{ \centering \PCHG, \PEXT, \PCHGr, \PEXTr }} & \raisebox{-0.4cm}{\parbox{6cm}{example of adding persistence to the transition; it is likewise for the rest }}  \\ \hline  	
		\end{tabular}
\end{table}

\subsection{Formal specification of \ourERT}

The logic-based reconstruction follows the usual approach as in earlier work \cite{Ar:Fr:er-99,Artale03,AK08is,Ongoma15}. 
Like \ERVT and \ERVTpp, \ourERT supports
timestamping for classes, attributes, and relationships and has both a textual and a graphical syntax, along with a model-theoretic semantics (as a temporal extension of the EER
semantics~\cite{Calvanese99}). The basic \ERVT was de facto extended with transition constraints (relation migration) in \cite{KA10} and with temporal attributes in \cite{OKM14,KO15} to result in \ERVTpp. This was extended with more mandatory vs optional transitions, more precise language features of EER in the formalisation thanks to the consolidation of separate extensions, and an evaluated notation for \ourERT.  The formal foundations of  
\ERVT has been proven to have a correct encoding of \ERVT schemas as knowledge base in \DLRUS (see \cite{Artale02,Artale03} for details). The extensions up to and including the features in \ourERT avail of that and are still expressible within \DLRUS, as shall be shown in the definition further below. 

Here, we first introduce formally the textual syntax of \ourERT and its mapping to the graphical syntax. This is followed by the \ourERT semantics and, finally, the mapping of \ourERT into \DLRUS. 

\begin{definition}[\ourERT Conceptual Data Model]\label{def:syntax}
  An \ourERT conceptual data model is a tuple: $\schema = (\LS, \rel,
  \att, \crd_A, \crd_R, \isa_C, \isa_R, \isa_U, \disj_C, 
   \disj_R, \cover, \as,\at, \ident, \mathcal{E})$, such that: \LS is
  a finite alphabet partitioned into the sets: \CS ({\em class}
  symbols), \AS ({\em attribute} symbols), \RS ({\em relationship}
  symbols), \US ({\em role} symbols), \DS ({\em domain} symbols),
  and \FS ({\em attribute role} symbols) which symobolises {\tt From} and {\tt To}, where
the set \CS of class symbols is partitioned into a set \CSS of
  {\em {\sc S}napshot classes} (marked with an {\sf
    S}), 
  a set \CSM of {\em {\sc M}ixed classes} (unmarked
  classes), 
  and a set \CST of {\em {\sc T}emporary classes} (marked with a {\sf
    T}). 
  A similar partition applies to the set \RS and to \AS, 
  and
\begin{compactenum}
\item $\att$ is a function that maps a class symbol in \CS to an
  \AS-labeled tuple over \DS, $\att(C)=\langle A_1:D_1,\ldots,A_h:D_h
  \rangle$.
\item $\rel$ is a function that maps a relationship symbol in \RS to
  an \US-labeled tuple over \CS, $\rel(R)=\langle
  U_1:C_1,\ldots,U_k:C_k \rangle$, and $k$ is the {\em arity} of $R$. 
To facilitate constraint specification, 
  we introduce the $\player$ and $\role$ functions, as follows: if $U_i: C_i \in \rel(R)$ then $\player(R,U_i) = C_i$ and $\role(R,C_i) = \{U_i\}$. The signature of the relationship is $\sigma_R \langle \US, \CS,\player,\role \rangle$, where for all $U_i \in \US$, $C_i \in \CS$, 
  if $\sharp U \geq \sharp C$, then for each $U_i, C_i, \rel(R)$, we have $\player(R,U_i) = C_i$ and $\role(R,C_i) = \{U_i\}$, and 
  if $\sharp U > \sharp C$, then there must be a more than one role played by the same class, i.e., at least once $\player(R,U_i) = C_i$, $\player(R,U_{i+1}) = C_i$ and $\role(R,C_i) = \{U_i,U_{i+1} \}$.
\item $\crd_R$ is a function $\CS\times\RS\times\US \mapsto \mathbb
  N\times(\mathbb N \cup\{\infty\})$ denoting cardinality
  constraints. We denote with $\cmin(C,R,U)$ and $\cmax(C,R,U)$ the
  first and second component of
  $\crd_R$. 
\item $\crd_A$ is a function $\CS\times\AS\times\FS \mapsto \mathbb
  N\times(\mathbb N \cup\{\infty\})$ denoting cardinality
  for attributes. We denote with $\cmin(C,A,F)$ and $\cmax(C,A,F)$ the
  first and second component of
  $\crd_A$. Note that $\crd_A(C,A,F)$ may be defined only if $(A,D)$ in $\att(C)$ for some $D \in \DS$.
\item $\isa_C$ is a binary relationship $\isa\subseteq
  (\CS\times\CS)$.  
\item $\isa_R$ is a binary relationship $\isa\subseteq
 (\RS\times\RS)$, which is restricted to relationships with the same arity 
 and  compatible signatures. 
\item $\isa_U$ is a binary relationship $\isa\subseteq (\US\times\US)$.  $\isa$ between roles of a relationship is restricted to relationships with 
 compatible signatures.
\item $\disj_C,\cover_C$ are binary relations over
  $(2^\CS\times\CS)$, describing disjointness
  and covering (`total'), respectively, over a group of $\isa_C$ that
  share the same superclass.  
\item $\disj_R$ are binary relations over
  $(2^\RS\times\RS)$, describing disjointness over a group of $\isa_R$ that
  share the same super-relation (if explicitly declared). 
\item $\ident$ is a function, $\ident:\CS\rightarrow\AS$, that maps a
  class symbol in $\CS$ to its identifier (`key') attribute and $A \in \AS$ is an attribute defined in $\att(C)$, i.e., $\ident(C)$ may be declared only if $(A,D) \in \att(C)$ for some $D \in \DS$.  
\item $\ES$ is the set of transition constraints, holding either between classes, between relationships, or between attributes, including 
$\chg$ and $\ext$ for classes, relationships and attributes for the past and $\CHG$ and $\EXT$ for the future, their mandatory counterparts ($\mchg$, $\mext$, $\MCHG$, $\MEXT$), quantitative change or extension in the past and future ($\qchg$, $\qext$, $\QCHG$, $\QEXT$), and $\frz$ for frozen attribute.\footnote{This list of transition constraints can be extended with `strong' or `persistent' versus reversible transition.} A particular constraint for a class (resp. relationship, attribute) is then denoted as, e.g., $\CHG_{C_i,C_j}$ (resp. $\CHGr_{R_i,R_j}$ and {\sc chgA}$_{A_i,A_j}$) where $i \neq j$, and listing the elements in the order of $_{source,target}$, and for a frozen attribute, the attribute that it applies to, as in \frz$_A$. Quantitative constraints are indicated in the superscript, adhering to the pattern $\qext^n_{C_i,C_j}$ (resp. for relationships and attributes) where $n$ is the unit of measure for the time interval.
\end{compactenum}
\end{definition}
Observe that EER typically does not contain relationship disjointness, nor subsumption on relationship components (roles), and only a few variants admit cardinality constraints on attributes that are permitted here as well. In contrast, since $\DLRUS$ does not have extensive support for identification and functional dependency support as its sibling language $\dlrifd$ has, $\ident$ over weak entity types can be declared in the syntax but will not have a corresponding model-theoretic semantics. While these features can be added to $\DLRUS$ since it is undecidable already anyway, we chose not to change the logic as such new machinery would distract from the key contribution. 

The mappings between the textual syntax and the graphical syntax of \ourERT is included in Table~\ref{tab:elements} for the elements and Table~\ref{tab:constraints} for the constraints.

\begin{figure}[h]
\centering
\includegraphics[width = 1.0\textwidth]{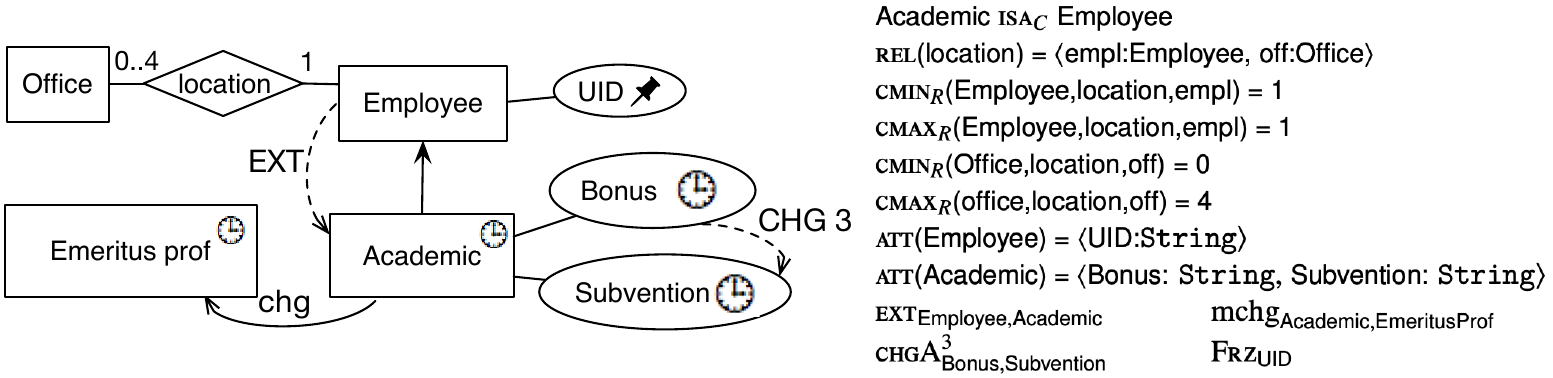}
\caption{A small partial \ourERT diagram (left) and corresponding textual notation (right); see text for explanation.} 
\label{fig:trendex}
\end{figure}

The \ourERT syntax is not designed for human usage as a way to create a conceptual data model, because the diagram notation is. The syntax is meant to facilitate the formalisation and as an exchange syntax to any other notation, such as possibly a coding style notation or a controlled natural language as in \cite{Keet17creol}, which were not preferred but not widely disliked options either \cite{KB17}. To illustrate the textual syntax nonetheless, consider the \ourERT diagram in Fig.~\ref{fig:trendex} and its textual notation. Key interesting aspects are that a bonus scheme for an academic may change into a subvention scheme (to top up their salary) after three years, and the transition constraints, whose arrows point in the direction whereto the entity will/does/did migrate. Specifically, each emeritus professor definitely used to be an academic employee and, optionally, an employee may become an academic. Further $\CST$ consists of {\sf\small Emeritus prof} and {\sf\small Academic} and $\AST$ consists of {\sf\small UID}, {\sf\small Bonus}, and {\sf\small Subvention}.

The set-based and model-theoretic semantics of the basic \ERVT has been described in \cite{Artale03,Artale07a,AK08dl}, which was updated in \cite{Ongoma15} for the \ERVTpp extensions notably temporalising attributes and refinements in syntax and formalisation\footnote{Notably that of the attribute set, more clearly distinguishing between object $o$, attribute $a$, and domain $d$.}, and again here with the \ourERT extensions. The key differences compared to \ERVTpp is the set of transition constraints $\ES$, of which a systematic full formalisation has been included here (not present in \cite{Ongoma15}) and as in \cite{Keet17creol} except for updates in \ourERT syntax, minor typos, and only those that were finally included in \ourERT. 

The \ourERT conceptual data modelling language adopts the snapshot representation of temporal conceptual data models~\cite{chomicki:toman:tl-book-98}. Specifically, following the
  snapshot paradigm, we have that ${\TSS}$ is a set of time points (also called chronons) 
  and $<$ is a binary precedence relation on ${\TSS}$, the flow of 
  time $\TS =\auf {\TSS},<\zu$ is assumed to be isomorphic to either
  $\auf\mathbb{Z},<\zu$ or $\auf\mathbb{N},<\zu$.\footnote{And so, standard
  relational databases can be regarded as the result of mapping a
  temporal database from time points in ${\TS}$ to atemporal
  constructors, with the same interpretation of constants and the same
  domain.} 
\begin{definition}[\ourERT Semantics]\label{er-sem}
  Let $\schema$ be an \ourERT conceptual data model. A {\em temporal database state}
  for the conceptual data model $\schema$ is a tuple
  $\B=(\TS,\Bint\Delta\cup\act,\Bintt\cdot)$, such that: $\Bint\Delta$
  is a nonempty set of abstract objects disjoint from $\act$;
  $\act=\bigcup_{D_i\in\DS}\acti$ is the set of basic domain values
  used in the schema $\schema$; and $\Bintt\cdot$ is a function that
  for each $t\in\TS$ maps:
     \begin{compactitem}
     \item Every basic domain symbol $D_i$ into a set
       $\Bintt{D_i}=\acti$.
     \item Every class $C$ to a set
       $\Bintt{C}\subseteq\Bint\Delta$---thus {\em objects} are
       instances of classes.
     \item Every relationship $R$ to a set $\Bintt{R}$ of \US-labeled
       tuples over $\Bint\Delta$---i.e. let $R$ be an n-ary
       relationship connecting the classes $C_1,\ldots,C_n$, $\rel(R)
       = \langle U_1:C_1,\ldots,U_n:C_n \rangle$, then, $r\in\Bintt{R}
       \to (r = \langle U_1:o_1,\ldots,U_n:o_n \rangle \land \forall
       i\in\set{1,\ldots,n}\per o_i \in\Bintt{C_i})$.  We adopt the
       convention: $\langle U_1:o_1,\ldots,U_n:o_n \rangle \equiv
       \langle o_1,\ldots,o_n \rangle$, when \US-labels are clear from
       the context.
     \item Every attribute $A$ to a set
       $\Bintt{A}\subseteq\Bint\Delta\times\act$, such that, for each
       $C\in\CS$, if $\att(C)=\langle A_1:D_1,\ldots,A_h:D_h \rangle$,
       then, $o\in\Bintt{C}\to (\forall i\in\set{1,\ldots,h}, \exists
       d_i\per a \in\Bintt{A_i} \land\forall
       d_i\per a_i \in\Bintt{A_i} \to d_i\in\acti)$.
     \end{compactitem}

     \noindent \B is said a {\em legal temporal database state} if it
     satisfies all of the constraints expressed in the schema,
     i.e. for each $t\in\TS$:

\begin{compactitem}
\item For each $C_1,C_2\in\CS$, if $C_1\isa_C C_2$, then,
  $\Bintt{C_1}\subseteq\Bintt{C_2}$.
\item For each $R_1,R_2\in\RS$, if $R_1\isa_R R_2$, then,
  $\Bintt{R_1}\subseteq\Bintt{R_2}$.
\item For each $U_1,U_2\in\RS$, if $U_1\isa_U U_2$, then,
  $\Bintt{U_1}\subseteq\Bintt{U_2}$.  
\item For each cardinality constraint $\crd_R(C,R,U)$, then:\\
  $o\in\Bintt{C}\to\cmin(C,R,U) \leq \#\{r\in \Bintt R \mid \player(r,U)=o \}
  \leq \cmax(C,R,U)$.
\item For each cardinality constraint $\crd_A(C,A,F)$, then:\\
  $o\in\Bintt{C}\to\cmin(C,A,F) \leq \#\{a\in \Bintt A \mid \player(a,{\tt From}) =o \}
  \leq \cmax(C,A,F)$. 
\item For $C,C_1,\ldots,C_n\in\CS$, if $\set{C_1,\ldots,C_n}\disj_C C$,
  then, \\ $\forall i\in\set{1,\ldots,n}\per C_i\isa_C C \land \forall
  j\in\set{1,\ldots,n},$ $j\neq i\per\Bintt{C_i}\cap\Bintt{C_j} =
  \emptyset$.\\ (Similar for $\set{R_1,\ldots,R_n}\disj_R R$)
\item For $C,C_1,\ldots,C_n\in\CS$, if $\set{C_1,\ldots,C_n}\cover C$,
  then, \\ $\forall i\in\set{1,\ldots,n}\per C_i\isa_C C \land
  \Bintt{C}=\bigcup_{i=1}^n\Bintt{C_i}$. 
\item For each snapshot class $C\in\CSS$, then,
  $o\!\in\!\Bintt{C}\to\forall t'\!\in\!\TS\per
  o\!\in\!\Bintx{C}{t'}$.
\item For each temporal class $C\in\CST$, then,
  $o\!\in\!\Bintt{C}\to\exists t'\!\neq\!t\per
  o\!\not\in\!\Bintx{C}{t'}$.
\item For each snapshot relationship $R\!\in\!\RSS$, then,
  $r\!\in\!\Bintt{R}\to\forall t'\!\in\!\TS\per
  r\!\in\!\Bintx{R}{t'}$.
\item For each temporal relationship $R\!\in\!\RST$, then,
  $r\!\in\!\Bintt{R}\to\exists t'\!\neq\! t\per
  r\!\not\in\!\Bintx{R}{t'}$.
\item For each snapshot attribute $A\!\in\!\ASS$, then,
  $a\!\in\!\Bintt{A}\to\forall t'\!\in\!\TS\per
  a\!\in\!\Bintx{A}{t'}$.
\item For each temporal attribute $A\!\in\!\AST$, then,
  $a\!\in\!\Bintt{A}\to\exists t'\!\neq\! t\per
  a\!\not\in\!\Bintx{A}{t'}$.
\item For each class $C\in\CS$, if $\att(C)=\langle
  A_1:D_1,\ldots,A_h:D_h \rangle$, and $\langle C,A_i\rangle\in\as$,
  then, $(o\in\Bintt{C}\land a_i \in\Bintt{A_i}) \to
  \forall t'\in\TS\per a_i \in\Bintx{A_i}{t'}$.
\item For each class $C\in\CS$, if $\att(C)=\langle
  A_1:D_1,\ldots,A_h:D_h \rangle$, and $\langle C,A_i\rangle\in\at$,
  then, $(o\in\Bintt{C}\land a_i\in\Bintt{A_i}) \to
  \exists t'\neq t\per a_i \not\in\Bintx{A_i}{t'}$.
\item For each $C\in\CS, A\in\AS$ such that $\ident(C)=A$, then, $A$ is
  a snapshot attribute---i.e. $ A \in\ASS$---and
  $\forall d\in\act\per \#\set{o\in\Bintt{C}\mid a \in\Bintt{A}}\leq 1$.
\item For the transition constraint in $\ES$:
\begin{compactitem}

\item 
$\EXT$, 
extension in the future, optional \\ $ o \in \Bintt{\mbox{\EXT}_{C_1,C_2}} \rightarrow ( o \in  \Bintt{{\tt C_1}}  \land o \notin \Bintt{{\tt C_2}} \land o \in \Bintx{{\tt C_2}}{t+1})$; 

\item 
$\MEXT$, 
Mandatory $\EXT$, \\ $o \in \Bintt{\mbox{\MEXT}_{C_1,C_2}} \rightarrow (o \in\Bintt{{\tt C_1}} \rightarrow  \exists t'>t.  o \in\Bintx{\mbox{\EXT}_{C_1,C_2}}{t'})$; 

\item 
$\ext$, 
extension in the past, optional\\ 
$ o \in \Bintt{\mbox{\ext}_{C_1,C_2}} \rightarrow (
 o \in \Bintt{{\tt C_1}}  \land 
 o \in \Bintt{{\tt C_2}} \land 
 o \notin \Bintx{{\tt C_2}}{t-1})$;

\item 
$\mext$, 
Mandatory $\ext$, past \\ $o \in \Bintt{\mbox{\mext}_{C_1,C_2}} \rightarrow (o \in\Bintt{{\tt C_1}} \rightarrow  \exists t'<t.  o \in\Bintx{\mbox{\EXT}_{C_1,C_2}}{t'})$; 

\item 
$\CHG$, 
change (evolution), future, optional \\
$ o \in \Bintt{\mbox{{\CHG}}_{C_1,C_2}} \rightarrow ( o \in  \Bintt{{\tt C_1}}  \land o \notin \Bintt{{\tt C_2}} \land o \in \Bintx{{\tt C_2}}{t+1} \land o \notin \Bintx{{\tt C_1}}{t+1})$; 

\item 
$\MCHG$, 
Mandatory change (evolution), future \\
 $o \in \Bintt{\mbox{\MCHG}_{C_1,C_2}} \rightarrow (o \in\Bintt{{\tt C_1}} \rightarrow  \exists t'>t.  o \in\Bintx{\mbox{\CHG}_{C_1,C_2}}{t'})$;  

\item 
$\chg$, 
past, optional \\
$ o \in \Bintt{\mbox{{\chg}}_{C_1,C_2}} \rightarrow ( 
o \notin\Bintt{{\tt C_1}} \land 
o \in \Bintt{{\tt C_2}} \land 
o \notin \Bintx{{\tt C_2}}{t-1} \land 
o \in \Bintx{{\tt C_1}}{t-1})$; 

\item 
$\mchg$, 
Mandatory change, past:\\
 $o \in \Bintt{\mbox{\mchg}_{C_1,C_2}} \rightarrow (o \in\Bintt{{\tt C_1}} \rightarrow  \exists t'<t.  o \in\Bintx{\mbox{\CHG}_{C_1,C_2}}{t'})$; 

\item 
$\PEXT$/$\PCHG$, 
Persistent extension or change;  persistence-part of the constraint, for classes (similar for relationships and attributes): \\
$o \in \Bintt{{\tt C_1}} \rightarrow \forall t' >t.  o \in \Bintx{{\tt C_1}}{t'}$;

\item 
$\QEXT$, 
Quantitative extension, future, optional, where here and in the following variants, $n \in \mathbb{Z}$ and $t+n \in \mathcal{T}_p$, and for $\ext n$ then: \\
$o \in \Bintt{\mbox{\QEXT}_{C_1,C_2}} \rightarrow \exists (t+n) > t. (o \in \Bintt{{\tt C_1}} \land o \notin \Bintt{{\tt C_2}} \land \Bintx{{\tt C_2}}{t+n} )$; 

\item 
$\MQEXT$, 
Quantitative extension, future, mandatory\\
 $o \in \Bintt{\mbox{\MQEXT}_{C_1,C_2}} \rightarrow (o \in\Bintt{{\tt C_1}} \rightarrow  \exists (t+n)>t.  o \in\Bintx{\mbox{\QEXT}_{C_1,C_2}}{t+n})$; 
 
\item 
$\qext$, 
Quantitative extension, past, optional \\ 
$o \in \Bintt{\mbox{\qext}_{C_1,C_2}} \rightarrow \exists (t-n) < t. (o \in \Bintx{{\tt C_1}}{t-n} \land o \in \Bintt{{\tt C_2}} \land o \notin \Bintx{{\tt C_2}}{t-n} )$; 

\item 
$\mqext$, 
Quantitative extension,  past, mandatory \\ 
$o \in \Bintt{\mbox{\mqext}_{C_1,C_2}} \rightarrow (o \in\Bintt{C_1} \rightarrow  \exists (t-n) < t.  o \in\Bintx{\mbox{\QEXT}_{C_1,C_2}}{t-n})$; 
 
\item  
$\QCHG$, 
Quantitative change, future, optional\\
$o \in \Bintt{\mbox{\QCHG}_{C_1,C_2}} \rightarrow \exists (t+n) > t. (o \in \Bintt{{\tt C_1}} \land o \notin \Bintt{{\tt C_2}} \land o \in \Bintx{{\tt C_2}}{t+n} \land o \notin \Bintx{{\tt C_1}}{t+n})$; 
 
\item 
$\MQCHG$, 
Quantitative change, future, mandatory\\  
$o \in \Bintt{\mbox{\MQCHG}_{C_1,C_2}} \rightarrow (o \in\Bintt{C_1} \rightarrow  \exists (t+n) > t.  o \in\Bintx{\mbox{\QCHG}_{C_1, C_2}}{t+n})$; 
 
\item 
$\qchg$, 
Quantitative change, past, optional\\ 
$o \in \Bintt{\mbox{\qchg}_{C_1,C_2}} \rightarrow \exists (t-n) < t. (o \in \Bintt{C_1} \land o \notin \Bintt{C_2} \land o \in \Bintx{C_2}{t-n} \land o \notin \Bintx{C_1}{t-n})$;

\item 
$\mqchg$,  
Quantitative change, past, mandatory\\ 
$o \in \Bintt{\mbox{\mqchg}_{C_1,C_2}} \rightarrow (o \in\Bintt{C_1} \rightarrow  \exists (t-n) < t.  o \in\Bintx{\mbox{\QCHG}_{C_2, C_1}}{t-n})$; 
 
\item 
$\EXTr$, 
extension for relationships, future, optional \\
 $ \langle o , o' \rangle \in \Bintt{\mbox{\EXTr}_{R_1,R_2}} \rightarrow (  
 \langle o , o' \rangle \in\Bintt{{\tt R_1}} \land 
 \langle o , o' \rangle \notin\Bintt{{\tt R_2}} \land 
  \langle o , o' \rangle \in\Bintx{{\tt R_2}}{t+1})$; 

\item 
$\MEXTr$, 
 extension for relationships, mandatory \\ 
 $ \langle o , o' \rangle \in \Bintt{\mbox{\MEXTr}_{R_1,R_2}} \rightarrow (\langle o , o' \rangle \in\Bintt{{\tt R_1}} \rightarrow  \exists t'>t. \langle o , o' \rangle \in\Bintx{\mbox{\EXTr}_{R_1,R_2}}{t'})$; 

\item 
$\extr$,  
 extension for relationships, past, optional \\  
 $ \langle o , o' \rangle \in \Bintt{\mbox{\extr}_{R_1,R_2}} \rightarrow (
 \langle o , o' \rangle \in\Bintt{{\tt R_1}} \land 
  \langle o , o' \rangle \in\Bintt{{\tt R_2}} \land 
 \langle o , o' \rangle \notin\Bintx{{\tt R_2}}{t-1})$. 

\item 
$\mextr$, 
 extension for relationships, past, mandatory \\ 
$ \langle o , o' \rangle \in \Bintt{\mbox{\mextr}_{R_1,R_2}} \rightarrow (\langle o , o' \rangle \in\Bintt{{\tt R_1}} \rightarrow  \exists t'<t. \langle o , o' \rangle \in\Bintx{\mbox{\extr}_{R_1,R_2}}{t'})$;
 
\item 
$\CHGr$, 
 change for relationships, future, optional \\
 $ \langle o , o' \rangle \in \Bintt{\mbox{\CHGr}_{R_1,R_2}} \rightarrow (  
 \langle o , o' \rangle \in\Bintt{{\tt R_1}} \land
\langle o , o' \rangle \in\Bintx{{\tt R_2}}{t+1} \land 
\langle o , o' \rangle \notin\Bintx{{\tt R_1}}{t+1})$; 
 
\item 
$\MCHGr$
change for relationships, future, mandatory \\
 $ \langle o , o' \rangle \in \Bintt{\mbox{\MCHGr}_{R_1,R_2}} \rightarrow (\langle o , o' \rangle \in\Bintt{{\tt R_1}} \rightarrow  \exists t'>t. \langle o , o' \rangle \in\Bintx{\mbox{\CHGr}_{R_1,R_2}}{t'})$;  
  
\item 
$\chgr$, 
change for relationships, past, optional\\
 $ \langle o , o' \rangle \in \Bintt{\mbox{\chgr}_{R_1,R_2}} \rightarrow (  
\langle o , o' \rangle \notin\Bintt{{\tt R_1}} \land
\langle o , o' \rangle \in\Bintt{{\tt R_2}} \land
\langle o , o' \rangle \in\Bintx{{\tt R_1}}{t-1} \land 
\langle o , o' \rangle \notin\Bintx{{\tt R_2}}{t-1})$; 
 
\item 
$\mchgr$, 
change for relationships, past, mandatory \\ 
$ \langle o , o' \rangle \in \Bintt{\mbox{\mchgr}_{R_1,R_2}} \rightarrow (\langle o , o' \rangle \in\Bintt{{\tt R_1}} \rightarrow  \exists t'<t. \langle o , o' \rangle \in\Bintx{\mbox{\CHGr}_{R_1,R_2}}{t'})$; 

\item 
$\frz$, 
``frozen'' attribute \\ 
$a \in \Bintt{\mbox{\frz}} \rightarrow \forall t' > t. a \in \Bintx{{\tt A}}{t'}$; 

\item 
$\achg$, 
attribute evolution, where $a$ is a binary relation between a class and a data type, \\
$a \in \Bintt{\mbox{\achg}_{A_1,A_2}} \rightarrow \exists t' > t. (a \in \Bintt{{\tt A_1}} \land a \notin \Bintt{{\tt A_2}} \land a \in \Bintx{{\tt A_2}}{t'} \land a \notin \Bintx{{\tt A_1}}{t'})$;  

\item 
$\Qachg$, 
Quantitative evolution of an attribute ($a$ is a binary relation between a class and a data type), \\
$a \in \Bintt{\mbox{\achg}_{A_1,A_2}} \rightarrow \exists (t+n) > t. (a \in \Bintt{{\tt A_1}} \land a \notin \Bintt{{\tt A_2}} \land a \in \Bintx{{\tt A_2}}{t+n} \land a \notin \Bintx{{\tt A_1}}{t+n})$ where $n \in \mathbb{Z}$;  

\end{compactitem}
\end{compactitem}
\end{definition}

Note that, while rather wordy and elaborate for formalisation, it is more versatile than a direct mapping into $\dlrus$ or any other temporal logic. Both modifying the icons or transition labels in the graphical notation and using only a well-defined fragment is easy to select and communicate. For the former, it would mean updating the table only, and for the latter only a one-liner on what is in $\Sigma$. It will also facilitate implementation, because it offers a text-based version of \ourERT. 
A straight-forward extension is to re-introduce all the attribute transitions, which follow the same pattern in formalisation and notation as the relationships.

\section{Evaluating modelling and understanding with \ourERT}
\label{sec:eval}

As no other modelling language supports all the temporal semantics incorporated in \ourERT, the ability to understand and model with \ourERT was evaluated by having students use this, and then indicate their own perceived comfort level with temporal concepts. The quality of the models they produced was assessed by recording the number of correct and incorrect uses of each temporal construct. Sect.~\ref{sec:e611design} describes the design of the experiments and Sect.~\ref{sec:e611rand} the results and discussion.

\subsection{Experiment design for E6-E11}
\label{sec:e611design}

Six  experiments were conducted with separate sets of participants, three with smaller groups of  postgraduate students (groups PG1 to PG3) and three with large and even larger cohorts of undergraduates (groups UG1 to UG3), of which E6 (with PG1), E7 (with PG2), and E8 (with UG1) took place in one year, E9 with UG2 and E10 with PG3 in another year (but such that UG1 participants could not have participated in PG3 later on), and E11 with UG3 after that. There are thus a number of elements and comparisons; while Table~\ref{tab:Experiments} lists the experiments in sequence and summarises them, Table~\ref{tab:names} lists the abbreviations and their descriptions, to facilitate keeping track of them all.
  
Overall, the experiments were designed to gauge whether \ourERT models could be correctly interpreted and constructed. Our hypotheses were that \ourERT models were understandable and that modellers would be able to design models effectively using \ourERT. Among the six experiments, there were slight variations based on results and insights obtained from the first round of experiments, which are described below.

\begin{table}[t]
\small
	\centering
	\caption{Experiment abbreviations, participant groups, models created and Multiple Choice Questions answered, in chronological order, totalling 1009 participants who created 246 models.}\label{tab:names}			
		\begin{tabular}{|p{0.35cm}|p{6cm}|p{4.6cm}|p{0.7cm}|}
\hline
{\bf No.} & {\bf Group that participated} & {\bf Set of Models} & {\bf MCQ}  \\ 	\hline \hline
E6 & PG1 (n=15); postgraduates (Masters or PhD) in Computer Science & M-PG1 (m=15) & - \\ \hline
E7 & PG2 (n=14); postgraduates, Masters in Data Science & M-PG2 (m=14) & q=14 \\ \hline
E8 & UG1 (n=177); 2nd year undergraduate students, database module & M-UG1 (m=43) & q=170  \\ \hline
E9 & UG2 (n=390); 2nd year undergraduate students, database module (different cohort from UG1) & M-UG2-C (m=38) with {\sc Chg} etc. and M-UG2-D (m=41) with {\sc Dev} etc. & q=373  \\ \hline
E10 & PG3 (n=11); postgraduates, Masters in IT & M-PG3 (m=11, with some as model fragments) & - \\ \hline								
E11 & UG3 (n=402); 2nd year undergraduate students, database module (different cohort from UG1 and UG2) & M-UG3-C (m=42) with {\sc Chg} etc. and M-UG3-D (m=41) with {\sc Dev} etc. & - \\ \hline
\end{tabular}	
\end{table}

\subsubsection{Experiment 6: understanding \ourERT diagrams}
Understanding was evaluated first with a small group of 15 postgraduate Computer Science students, PG1, as part of a study on modelling preferences \cite{KB17}. The single paragraph in Sect.~\ref{TrendSpec}, supplemented by a small example, introduced them to \ourERT. They were then given a \ourERT model with entity and relationship names such as ``P'' and ``Q'', and asked to interpret its five temporal components \cite{KB17}. Submissions were independently assessed by two of the authors on a scale from 1 (not understanding) to 5 (perfect understanding), compared, discussed, and harmonised. 

\subsubsection{Experiment 7 and 8: understand and create \ourERT diagrams}
\begin{figure}[t]
\centering
\includegraphics[width = 1.0\textwidth]{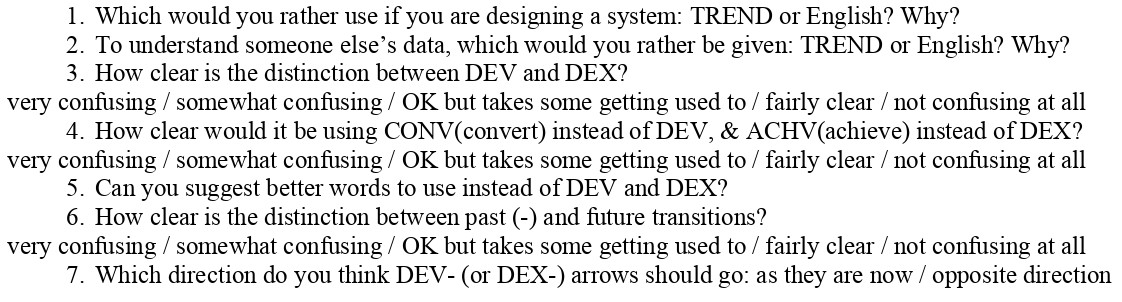}
\caption{Multiple Choice Questions posed to students in the UG1 group (E8). It is the same for UG2 in E9, except question 4, which was 
``How clear would it be using {\sc Tran} (transform) instead of {\sc Chg}, \& {\sc Also} instead of {\sc Ext}?'' for M-UG2-C and ``How clear would it be using {\sc Tran} (transform) instead of {\sc Dev} , \& {\sc Also} instead of {\sc Dex} ?'' for M-UG2-D.}
\label{fig:MCQs}
\end{figure}

The second and third experiment, with PG2 and UG1, expanded understanding to also construction of \ourERT models. For PG2, E7 was conducted during a lecture on data modelling to postgraduate Data Science students, who ranged from novice to proficient modellers. This lecture illustrated \ourERT using 3 small examples, along with a summary of the notation  showing its temporal constructs and a description of each (Fig.~\ref{fig:tibtable} in the Appendix). Fourteen students participated voluntarily, which required describing the temporal aspects of two small models, drawing a \ourERT model for a Loans Company (described in one paragraph), and answering seven multiple choice questions, which are shown in Fig.~\ref{fig:MCQs}. They were given 40 minutes to complete this in the presence of one of the authors. Descriptions of the given models were evaluated by recording understanding of each temporal construct as complete and correct, correct but incomplete, missing or wrong. This method of recording was also used in assessing their \ourERT models; since the brief was explicit, correctness was established by comparing each against the expected model. Finally, answers to the questions in Fig.~\ref{fig:MCQs} were recorded along with subjects' 
home language and stated level of comfort with data modelling.

The aims and set-up of the large-scale E8 for the UG1 group will be described in the subsection below, for readability.

After analysis of Experiments 6-8, three more were designed. Experiment 9 (E9) with UG2 had as hypothesis that more explanation and different labelling of the transition constraints---based on suggestions from the UG1 students in E8---would result in better models and better understanding (details in subsection below). Experiment 10 (E10) with PG3 sought to exclude the possible effect that the students would not know enough of the subject domain and use the temporal constraints possibly incorrectly (or correctly for incorrect understanding) because of it and to assess understanding at a deeper level. Its outcomes motivated Experiment 11 (E11), which ran like E9, but with a more constrained modelling task (also described in the next subsection). The  
main aim to verify that \ourERT modelling and understanding outcomes are attributable to the language indeed, not the subject domain understanding or the lack thereof.

\subsubsection{Experiment 10: constraining the modelling task}
PG3 had carefully worded individual statements from the controlled natural language for \ourERT \cite{Keet17creol} of a universe of discourse that had to be modelled, one for each key construct of \ourERT. Their understanding was tested not through describing it but asking them how that temporal constraint might then be implemented in a database. All 11 students enrolled in the Masters in IT conversion course on Databases participated, since it was part of an assignment. The answers were evaluated manually by one of the authors. The  model snippets were coded with a Yes/Partial/No and aggregates were computed. The answers of the understanding question were analysed and summed up in keywords and coded as 0=no answer, 1=an off-topic generic answer, 2=wrong answer, and 3=something plausible. The latter category is flexible because there are no algorithms and agreed upon transformations from temporal conceptual data models to a database and therefore multiple options to achieve the same are possible. 

\subsubsection{Experiments E8, E9, and E11: the large-scale experiments on understanding and modelling} 
 The three large-scale experiments were conducted as part of the first data modelling assignment in the Second Year \emph{Introduction to Databases} module at the University of Cape Town (UCT). E8 involved the first test a large group, UG1, which aimed at determining which temporal aspects were most commonly used, eliciting many opinions on alternative notations and perceived understandability, and testing our initial hypotheses on a larger sample of models constructed by novices. E9 with UG2 provided an even larger sample for testing our two hypotheses that \ourERT models were understandable and novices able to design models using it effectively. It also explored whether a modified notation for transitions, and an expanded explanation of \ourERT, would lead to fewer errors and/or greater usage of temporal constructs by novices. Finally, E11 with UG3 ran like E9 except that the domain content to model was more constrained to allow for grading against a gold standard (known correct answer) by availing of its controlled natural language \cite{Keet17creol} to generate the description of the intended model and therewith also eliminating the possibility that students would not know enough of the domain. 

Run in different years, the UG1 and UG2 cohorts were asked to design a \ourERT model of UCT showing ``as many temporal characteristics and transitions as possible'', and to describe in English what their model shows. Students worked in groups of five to encourage debate and reflection. They were also required to individually answer the 7 multiple choice questions shown in Fig.~\ref{fig:MCQs}. 

\begin{figure}[h]
\centering
\includegraphics[width = 1.0\textwidth]{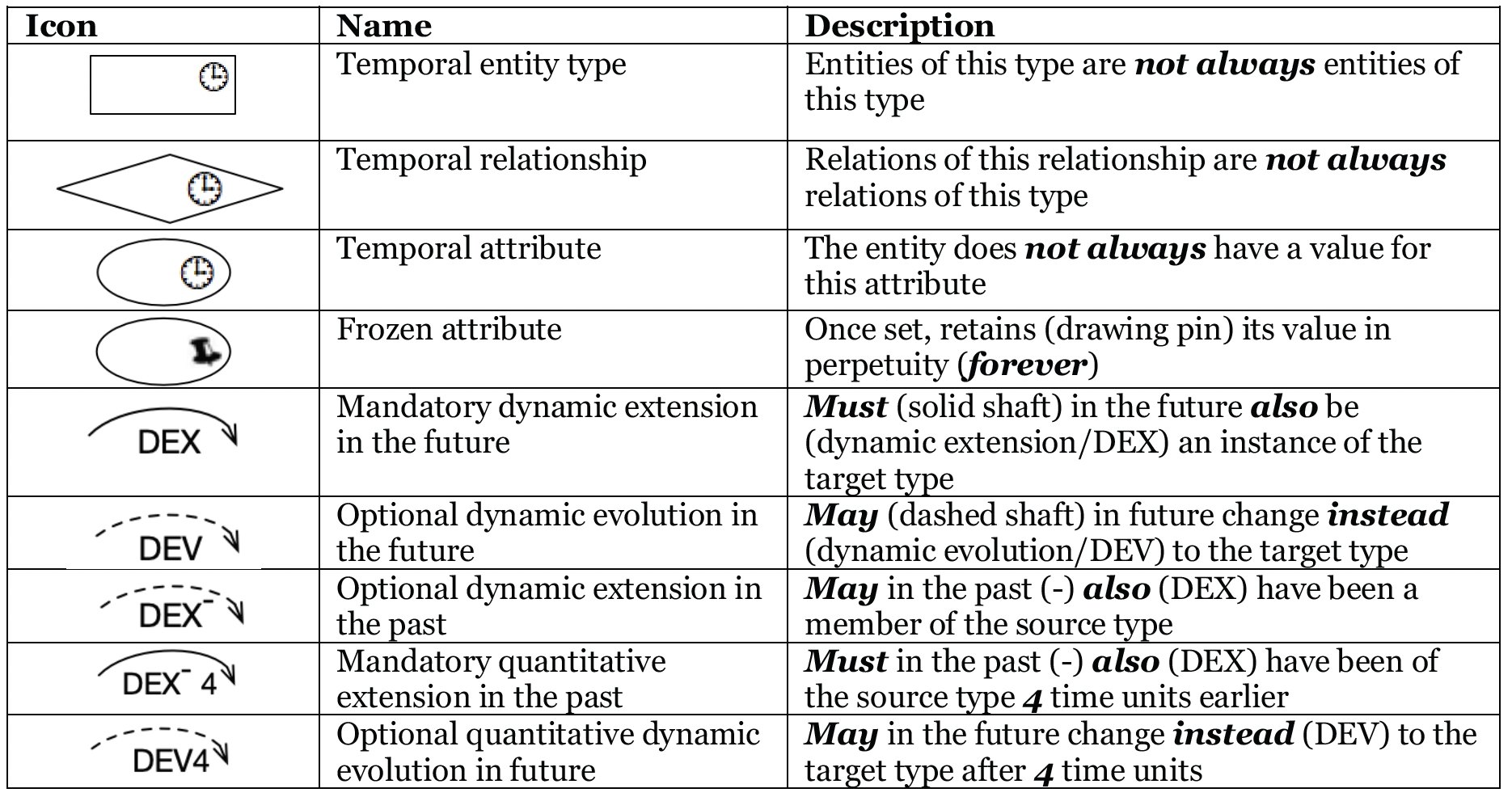}
\caption{Summary table of the base temporal notations of \ourERT as it was presented to the participants in the validation experiment, and likewise suitably modified for the {\sc Chg} and{\sc Ext variants} in later experiments.}
\label{fig:tibtable}
\end{figure}

The \ourERT extensions to ER modelling were introduced to the UG1 group in a 3-page document that combined the same paragraph (see Section~\ref{sec:TrendSpec}) used initially, with the summary table from \cite{KB17} (alike  Fig.~\ref{fig:tibtable}) and 3 small examples from the second experiment. For E9 and E11, this outline was extended to include a better explanation of the new concepts, a graphical depiction of temporal concepts, and a sample of controlled natural language sentences generated from one of the examples. 

\begin{figure}[h]
\centering
\includegraphics[width=0.85\textwidth]{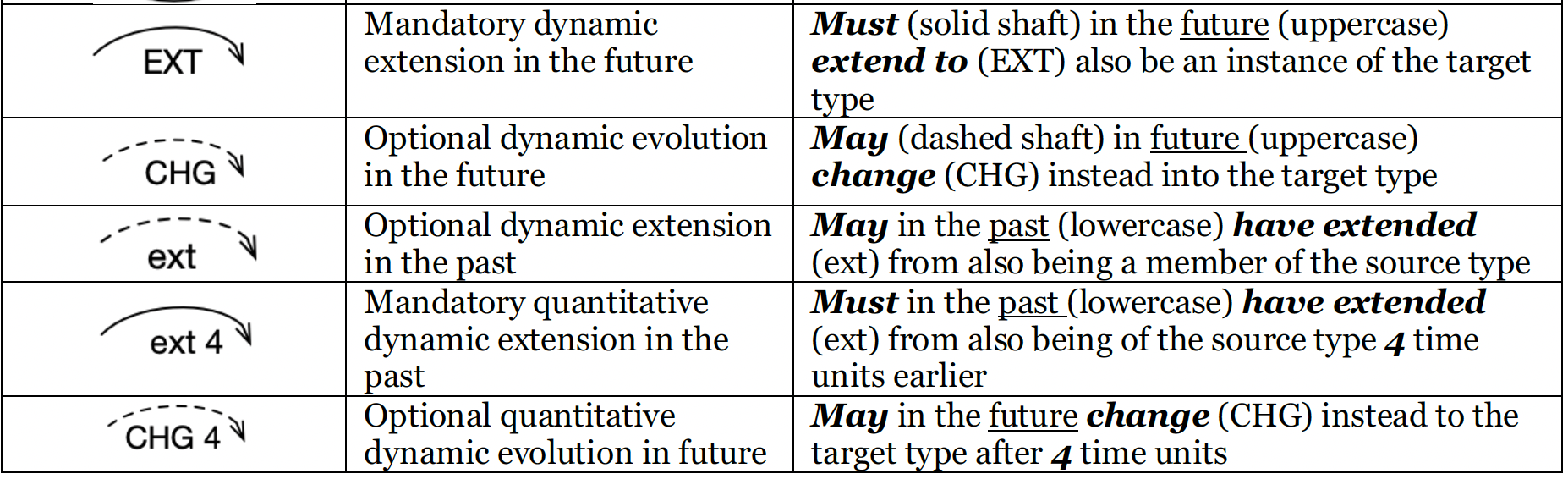}
\caption{Summary table of the revised temporal notation for the transition constraints (and corresponding wording in the explanation) for \ourERT, which was provided to the applicable subset of the experiments' participants from E9 onwards.}
\label{fig:tibtableCHG}
\end{figure}

\begin{figure}[h]
\centering
\includegraphics[width = 0.75\textwidth]{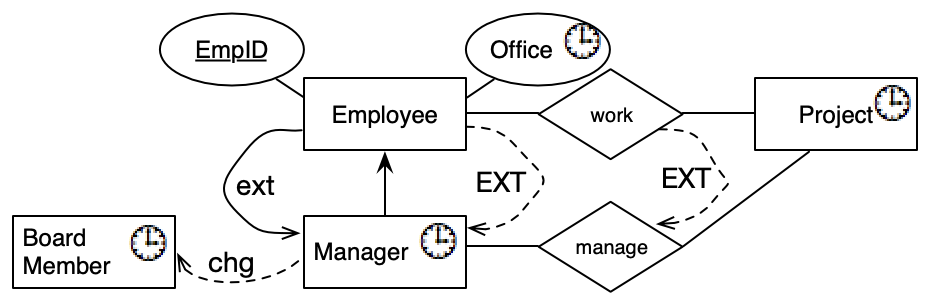}
\caption{One of the sample diagrams provided in the \ourERT outline that was provided to the applicable subset of the experiments' participants from E9 onwards.}
\label{fig:tibCHGex}
\end{figure}

The set-up of E9 was extended cf. E8 since the class had doubled in size. For E9, the UG2 group was divided in two, with one half given the original \ourERT, and the other half a new version in which {\sc Dev} and {\sc Dex} were replaced by {\sc Chg} (change) and {\sc Ext} (extend) respectively, and in which past and future transitions were distinguished using lower- or uppercase, rather than 
minus, (as indicated in Fig.~\ref{fig:tibtableCHG}) and 3 small examples from the second experiment, one of which is shown in Fig.~\ref{fig:tibCHGex}. For E8 and E9, there were thus 3 separate sets of students: UG1 generating a set of models \textbf{M-UG1}, who were given the \emph{original} \ourERT notation and \emph{short} \ourERT outline; UG2-C in the second cohort generating models in a set labelled \textbf{M-UG2-C}, who were given the \emph{new} \ourERT notation (\underline{C}hg etc.) and \emph{extended} outline; and UG2-D that created the set of models labelled \textbf{M-UG2-D}, who were given the \emph{original} \ourERT notation (\underline{D}ev etc.) but the \emph{extended} \ourERT description. There were 43, 41 and 38 models in sets M-UG1, M-UG2-D and M-UG2-C, respectively.

All 122 models were codified in an Excel spreadsheet. For each temporal construct, the number of times this appeared in each model was recorded, as well as the number of times incorrectly and correctly used. The total number of entities and relationships was kept as a rough indicator of model size against which to compare the number of temporal components. Choice of optional/mandatory, and direction of transitions in the past, were both encoded as Y for always correct, N for never correct, P for partly correct, and blank for never appearing in the model. In order not to penalise potential misunderstanding of the regulations at UCT, only definite errors were recorded as incorrect throughout. Thus, for example, specifying Name as a frozen attribute of Course was considered a reasonable, albeit false, assumption, and not counted as a modelling error.

Answers to the 7 multiple choice questions were also entered in Excel. 
As some students omitted these, there were 170, 195 and 178 responses from sets M-UG1, M-UG2-D and M-UG2-C respectively. 

Last, we considered the possibility that students may understand the temporal conceptual modelling but know too little, or 
be too inexperienced, to design a model from scratch on some given domain. For instance, they may understand {\sc Dev/Dex} but not the university's rules on course registrations and exclusion policies and a mistake about that in the model may be difficult to attribute to either the language or the domain understanding. To double-check this, we set up the third and final large-scale experiment,  E11, with the 2022 class of Introduction to Databases (UG3 as group) that was of similar size as UG2 of 2021. Two of the authors created a \ourERT diagram in the tourism domain and then used the controlled natural language of \cite{Keet17creol} to generate natural language sentences, such as ``Each traveller must evolve to a previous-customer ceasing to be a traveller.'' and ``Once the value for arrival time is set, it cannot change anymore.''. The text and the model answer are included in Appendix A4.

We again took advantage of the large group and split them up into two (named This and That for the students), 
where
This was given {\sc Chg/Ext} (whose set of created models are labelled \textbf{M-UG3-C} henceforth) and That was given the original {\sc Dev/Dex} (labelled \textbf{M-UG3-D}) to also assess whether the labelling affected the modelling in a more controlled setting. Both groups were given the same set of sentences and provided with the same \ourERT explanations as in E9. Groups of five students were created beforehand in the same manner as in E9. We did not include the 7 multiple choice questions\footnote{A quiz was added on the CMS to gain feedback on groupwork, but for a purpose unrelated to the \ourERT evaluation.}. The submitted answers were mostly marked by the tutors, following the marking rubric set by 
two
of the authors and the model answer of the diagram that the sentences were created from. There were 12 `buckets' for 1 mark each, whose total was divided by 2 for the overall mark for that subquestion of the assignment. Buckets include, among others, `Traveller, Previous Customer and VIP Customer have a clock' (yes:1, no:0), `any one ``past'' transition: VIPcust to PrevCust/Traveller; Took(flight) to Pays(flight)', and `all 3 above correct'. 

The marks were tabulated in a spreadsheet for analysis.

\subsection{Results and discussion}
\label{sec:e611rand}

We first describe the small-scale experiments E6, E7, and E10 that have a qualitative flavour, and subsequently the large-scale experiments E8, E9, and E11, which have a distinctly quantitative angle to the assessment of \ourERT.  

\subsubsection{Small-scale experiments}

The 14 Data Science students (PG2 in E7) clearly preferred using \ourERT to English. The only one who preferred specifications given in English instead, was also the only one ``not at all’’ comfortable with modelling and gave this as the reason for their choice. Only one of those ``fairly comfortable’’ with modelling preferred English to \ourERT for designing, saying it was ``easier to describe something in language (natural) than re-interpret visually’’. Altogether 4 preferred English and 10 preferred \ourERT for creating models. Reasons for preferring \ourERT were ``compact’’, ``quicker’’, ``easier’’, ``shows context’’, ``get overview’’, ``easier for brainstorming’’ and ``ensures you think critically”.  Difficulty with transitions were expressed only by 2 (who found this ``somewhat confusing’’), and only 1 felt transitions in the past should use the reverse direction. Their responses are summarised in Table~\ref{fig:migrateOK} and Table~\ref{fig:labelPref} further below, in the last column for easy comparison with the large scale experiments. Students' indication of familiarity with modelling showed a positive correlation with their choice of \ourERT or English for interpreting and creating models (0.50 and 0.60 respectively) but no correlation with their responses to questions about transitions. 

The Data Science students were given 2 \ourERT models and asked to explain in detail their temporal aspects. There was only one case of any misunderstanding, when a temporal attribute was given as ``can change’’. However, 3/4 of their explanations were imprecise, stating e.g. ``is a temporal attribute’’, explaining transitions as ``may/must become’’ without saying `additionally' or `instead', etc.

For the PG3 students in E10, eleven students submitted their assignments and they all attempted the third question on temporal aspects. Of the five modelling questions, five students answered more than half of them correctly, but none had all of them correct. Combining `yes' and `partial', 9/11 scored $\geq 3/5$ overall. Per question, the frozen attribute (Q3) was answered correctly most often (n=8) and then Q1 (optional DEV/CHG on entity types) and Q4 (optional DEX/EXT on entity types) with n=6. Curiously, Q5, a simple temporal entity type (i.e., with a clock icon), caused the most problems, with only 2 correct answers and 7 incorrect. 

As to the questions that tested deeper understanding of temporal information, and noting that temporal databases is part of the study material, it indicated little understanding of the subsumption and the transition constraint that had to be converted to a plausible implementation in the database. Three did not answer the question at all, 2 were off-topic, 3 were wrong, and only the remaining 3 had a plausible database solution. An example of a a plausible solution proposed was to add a column for {\tt Previous title} and a column for an attribute with datatype {\tt Date} that records the {\tt Promotion date}, and of a wrong answer was to ``add NOT NULL query''. The majority of the answers suggest that the modelling answer to the first question is an outcome of a novice still at the reproducibility or limited understanding level of temporal modelling.

\subsubsection{Large-scale experiments: E8, E9, and E11} 

\begin{figure}[h]
\centering
\includegraphics[width = 1.0\textwidth]{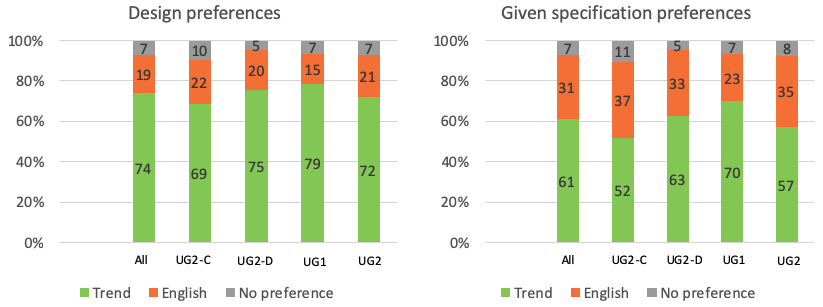}
\caption{Preference for English or \ourERT usage by the participants of E8 and E9.}
\label{fig:langPrefs}
\end{figure}

Of the total of 567 students (i.e., UG1+UG2) participating in experiments E8 and E9, 74\% preferred \ourERT and only 19\% preferred English when creating models (7\% had no preference), while 61\% preferred \ourERT and 31\% preferred English for understanding others’ model (8\% indicated no preference). The proportion preferring \ourERT to English for both purposes was higher in the UG1 group (in E8) than in the UG2 group (in E9), despite the extended explanation given to the latter, as shown in Fig.~\ref{fig:langPrefs}.

\begin{figure}[h]
\centering
\includegraphics[width =0.8\textwidth]{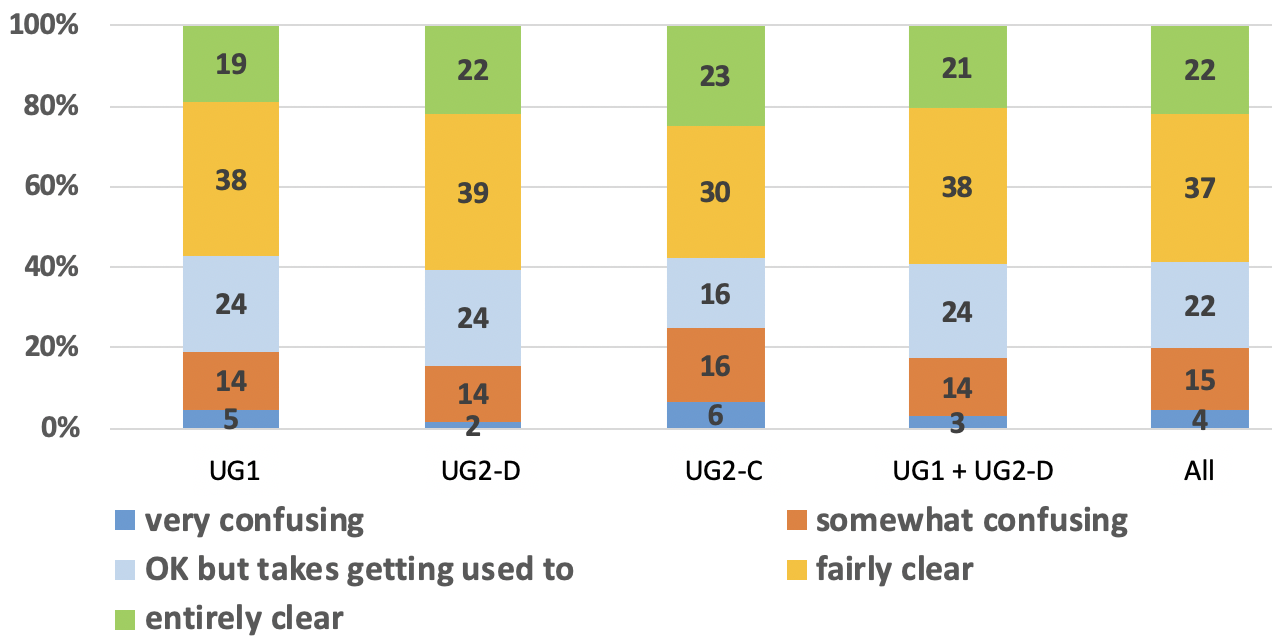}
\vspace{-1ex}
\caption{Self-reported rating of own ease/difficulty understanding transitions in the past (1 worst to 5 best).}
\label{fig:pastOK}
\end{figure}

While the clear preference for \ourERT is encouraging, fewer preferring \ourERT for interpreting models may mean some lack confidence in understanding it. Since none of these students were given either a \ourERT model or English description to interpret, however, it is probably more indicative of their failure to grasp the ambiguity and complexity of English specifications. Perceived understanding of \ourERT is important, but subsequent analysis revealed that this did not correlate well with their dynamic transition usage. Finally, among these 567 students, 22\% felt transitions in the past should be given in the opposite direction; 69\% agreed that the direction used was correct as is, and 9\% failed to answer. Their other responses on label preferences and understanding of transitions are summarised in Table~\ref{fig:migrateOK}, Table~\ref{fig:labelPref}, and Fig.~\ref{fig:pastOK}.

\begin{table}[t]
\small
	\centering
	\caption{Self-reported rating of own ease/difficulty understanding transitions, in percentages (question in E7, E8, and E9). 
	}\label{fig:migrateOK}			
		\begin{tabular}{|p{5cm}|p{1cm}|p{1.3cm}|p{1.3cm}|p{1.3cm}|p{1cm}|}
\hline
\textbf{Participant group $\rightarrow$ \newline
Answer option $\downarrow$} & \textbf{UG1} (n=164) & \textbf{UG2-D} (n=193) & \textbf{UG2-C} (n=178) &  \textbf{UG1 + UG2-D} (n=357) & \textbf{PG2} (n=14) \\ \hline \hline
{\sc Dev/Dex}  very confusing& 1  &0 & &0 & 0 \\ \hline
{\sc Dev/Dex} somewhat confusing  & 8  & 10& &9 &14 \\ \hline
{\sc Dev/Dex}  OK- takes getting used to  & 21  & \textbf{38} & &30 &\textbf{43} \\ \hline 
{\sc Dev/Dex} fairly clear & \textbf{36}  & \textbf{39} & & \textbf{38}&29 \\ \hline 
{\sc Dev/Dex} not confusing at all & \textbf{34}  &13 & &23 & 14\\ \hline
{\sc Chg/Ext} very confusing &  & &2 & & \\ \hline
{\sc Chg/Ext} somewhat confusing   &  & &10 & & \\ \hline
{\sc Chg/Ext} OK- takes getting used to &  & & 29& & \\ \hline 
{\sc Chg/Ext} fairly clear &  & & \textbf{42} & & \\ \hline
{\sc Chg/Ext}  not confusing at all &  & & 17& & \\ \hline 
\end{tabular}	
\end{table}

\begin{table}[h]
\small
	\centering
	\caption{Transition label preferences (in percent) among those who rated both given and alternate labels (question in E7, E8, and E9).}\label{fig:labelPref}			
		\begin{tabular}{|p{6.1cm}|p{1.2cm}|p{1.5cm}|p{1.5cm}|p{1cm}|}
\hline
\textbf{Participant group $\rightarrow$ \newline
Answer option $\downarrow$} & \textbf{UG1} (n=164) & \textbf{UG2-D} (n=193) & \textbf{UG2-C} (n=178) &  \textbf{PG2} (n=14)  \\ \hline \hline
Prefer {\sc Dev/Dex} to {\sc Conv/Achv} & \textbf{57} & & & 36 \\ \hline
Prefer {\sc Conv/Achv} to {\sc Dev/Dex}  & 18 & & & 36 \\ \hline
Rate {\sc Dev/Dex} and {\sc Conv/Achv} equally  & 26 & & & 28 \\ \hline
Prefer {\sc Dev/Dex} to {\sc Tran/Also}    & & \textbf{39} & & \\ \hline
Prefer {\sc Tran/Also} to {\sc Dev/Dex}     & & 40 & & \\ \hline
Rate {\sc Dev/Dex} and {\sc Tran/Also} equally     & & 21 & & \\ \hline
Prefer {\sc Chg/Ext} to {\sc Tran/Also}       & & & \textbf{45} & \\ \hline
Prefer {\sc Tran/Also} to {\sc Chg/Ext}        & & & \textbf{34} & \\ \hline
Rate {\sc Chg/Ext} and {\sc Tran/Also} equally   & &  & 21 & \\ \hline 
\end{tabular}	
\end{table}

\subsubsection{How well is \ourERT used by novice modellers?}

The Loans example given to the 14 Data Science students (PG2 in E7) 
clearly distinguished {\sf Current loans} and {\sf Historical loans}. Of the 11 who had these as 2 separate entities, 9 used transitions between these correctly (5 had {\sc Dev}, 2 had {\sc Dev-} and 2 had both), but 2 had {\sc Dex} instead. None drew the arrow in the wrong direction, and choice of optional or mandatory transition was always correct. Exactly half had attributes of {\sf Historical loans} as frozen. Half used temporal attributes wrongly, clearly under the misapprehension that this meant value changes over time. The two uses considered reasonable as evidence of understanding temporal attributes, were {\sf customer salary} and {\sf loan balance}. Attributes were never wrongly made frozen, but 8 left {\sf CustomerID} as snapshot. There was no correlation between student’s indicated comfort level with modelling and the quality of their \ourERT models.

\begin{figure}[h]
\centering
\includegraphics[width = 1.0\textwidth]{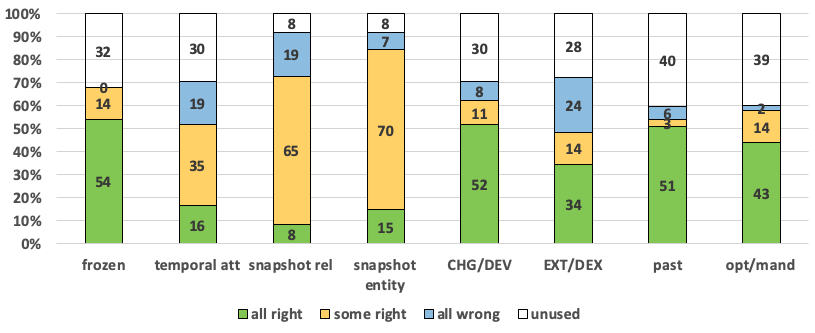}
\caption{Accuracy of usage by students new to data modelling, in all 122 models of E8 and E9 combined.}
\label{fig:constructYPN}
\end{figure}

\begin{figure}[h]
\centering
\includegraphics[width = 0.7\textwidth]{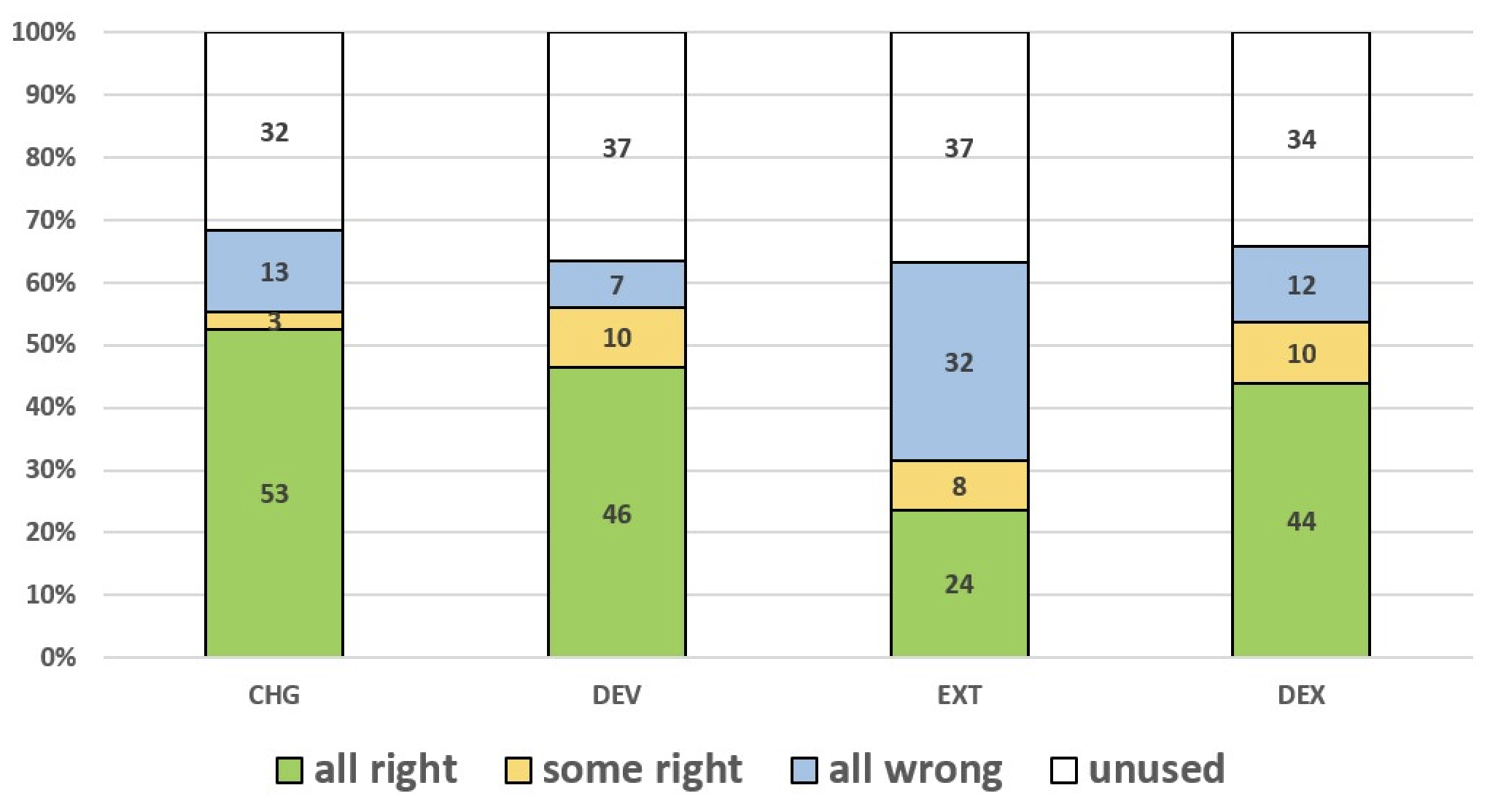}
\caption{Accuracy of transition usage by students new to data modelling, in the 79 models in M-UG2.}
\label{fig:migrateYPN}
\end{figure}

The large-scale studies,  involving novice modellers, revealed that the same misunderstanding of temporal attributes was common, and that dynamic extension was less well used than dynamic evolution, as shown in Table~\ref{fig:useOK}, Table~\ref{fig:prevalence}, Fig.~\ref{fig:constructYPN} and Fig.~\ref{fig:migrateYPN}.  
Frozen attributes, distinction between optional and mandatory transition, and the direction of the arrow for transitions in the past, were all correctly used in most models. Temporal entities were far more common than temporal relationships, and many models had temporal entities participating in atemporal relationships. Different labels for transitions did not impact accuracy of usage, and the improved explanation of \ourERT in the final experiment did not lead to better models. It should be noted, however, that unlike its predecessor, E9 involved students who had spent only one month on campus before having to learn Computer Science wholly online, due to the Covid-19 pandemic; so it is possible that they were less prepared than the students in UG1 of E8. Also, they were taught EER by a different (junior) lecturer unfamiliar with temporal conceptual models, which may have affected modelling with \ourERT which may have negatively affected modelling with \ourERT. 

\subsubsection{How well is each aspect used by novices?}
From the model descriptions required of students, it was clear that the concept of a frozen attribute was well understood. However, while attributes were rarely incorrectly marked as frozen, there were several models where only a subset of the frozen attributes were indicated on the diagram. The issue of whether or not primary keys should be shown as frozen or not was discussed on the class online forum, and remains unclear. It soon became evident that the need to indicate frozen attributes would alert modellers to their lack of knowledge about an organisation: e.g., students did not know which attributes would require a new course to be created at UCT were they to change, so NQF (South Africa's National Qualifications Framework) level, NQF credits, course name, etc. were considered frozen by some but not others. As regards the notation, the drawing pin was the only \ourERT symbol that modellers struggled to depict, albeit in very few cases.

The majority of modellers misunderstood temporal attributes as being those properties which can change over time, rather than those which, once set, may not always have a value. This was also borne out in their model descriptions, where this misconception was explicitly stated. Interestingly, while many of those models thus had every attribute contain either a pin or a clock, it is unclear what semantics the others associated with the attributes that had neither and were thus technically mixed attributes. On a more positive note from the perspective of the experiment, over 20 valid temporal attributes were depicted altogether, as a result of students adding attributes such as {\sf registration-hold} (due to unpaid fees) and {\sf prerequisite-subminimum},  and more typical or expected frozen attributes such as {\sf YearCompleted}, as well as declaring identifier attributes, such as {\sf StudentNumber}, frozen.

Most uses of dynamic evolution involved part (or all!) of the academic pathway from {\sf Applicant} to {\sf Undergraduate} ... {\sf  PhDstudent} to {\sf Lecturer} ... {\sf Professor}. Dynamic evolution for relationships was far less common; those that used this typically depicted this for student:course relationships, e.g., {\sf registers} becoming {\sf completes}, or {\sf planned} becoming {\sf enrols}, and from {\sf takes} a course to {\sf completes} or {\sf teaches} a course. Almost as many models included this evolution incorrectly by placing the transition on relationships of different types, such as from {\sf  Student takes Course} to {\sf Student achieves Result}. Valid dynamic extension usage was less common and limited to entity types; examples typically being {\sf Student} becomes {\sf Tutor} or the like, {\sf Lecturer} becomes {\sf Advisor} or the like, and {\sf registers} extends to {\sf completes}. In addition, some models had transitions connecting items of different kinds (e.g., from relationship to entity); but these were typically found where non-temporal aspects of models were also used incorrectly.  
An example of an optional quantitative transition constraint was a {\sc Dev 1} between a {\sf MAM1000WStudent} and {\sf MAM2(2LA)Student}, i.e., students optionally proceeding to second year mathematics in the next year, and an optional {\sc Dex 0.5} from {\sf Undergraduate} to {\sf Tutor}. 

\begin{table}[h]
\small
	\centering
	\caption{Temporal construct usage accuracy across all models in each set of models in E8 and E9.}\label{fig:useOK}		
		\begin{tabular}{|p{4.9cm}|p{1.1cm}|p{1.3cm}|p{1.3cm}|p{1.2cm}|p{1.1cm}|}
\hline
\textbf{Participant group $\rightarrow$ \newline
Transition Correctness $\downarrow$} & \textbf{M-UG1} (m=43) & \textbf{M-UG2-D} (m=41) & \textbf{M-UG2-C} (m=38) &  \textbf{M-UG2} (m=79) &  \textbf{All} (m=122)  \\ \hline \hline
Overall \% of temporal constructs correctly used (across all models) & 71 & 70 & 70 & 70 &70 \\ \hline
Overall \% {\sc Dev} transitions correct  &85 &81 & & \multirow{2}{*}{78} & \multirow{2}{*}{\textit{82}}\\ \cline{1-4}
Overall \% {\sc Chg} transitions correct& & & 75& & \\ \hline
Overall \% {\sc Dex} transitions correct&62 &70 & & \multirow{2}{*}{60} & \multirow{2}{*}{61}\\ \cline{1-4}
Overall \% {\sc Ext} transitions correct& & & 51& & \\ \hline
Overall \% {\sc Dev/Dex} transitions correct& 72& \textbf{76}& & \multirow{2}{*}{69} & \multirow{2}{*}{71}\\ \cline{1-4}
Overall \% {\sc Chg/Ext} transitions correct& & & \textbf{62} & & \\ \hline
Models with correct {\sc Dev/Chg} uses $>$ correct {\sc Dex/Ext} (in \%) &40 &34 &37 &35 &37 \\ \hline
Models with correct {\sc Dex/Ext} uses $>$ correct {\sc Dev/Chg} (in \%)&47 &27 &26 & 27& 34\\ \hline
Models with {\sc Dev/Chg} errors $>$ {\sc Dex/Ext} errors (in \%)&9 &10 &8 &9 &9 \\ \hline
Models with {\sc Dex/Ext} errors $>$ {\sc Dev/Chg} errors (in \%)& 30& 17& 29&23 &25 \\ \hline
\end{tabular}	
\end{table}

\begin{table}[h]
\small
	\centering
	\caption{Percentage of models using each construct (\% using correctly at least once in brackets), in Experiments E7-E9.}\label{fig:prevalence}		
		\begin{tabular}{|p{3.25cm}|p{1.1cm}|p{1.4cm}|p{1.4cm}|p{1.1cm}|p{1.15cm}|p{1.1cm}|}
\hline
\textbf{Participant group $\rightarrow$ \newline
Construct Correctness $\downarrow$} & \textbf{M-UG1} (m=43) & \textbf{M-UG2-D} (m=41) & \textbf{M-UG2-C} (m=38) &  \textbf{M-UG2} (m=79) &  \textbf{All} (m=122) &  \textbf{M-PG2} (m=14) \\ \hline \hline
Temporal entity types & \textbf{88} & 76 & 84 & 80 & 84 & 36\\ \hline
Temporal relationships & 37 & 34 & \textbf{39} & 38 & 38 &43 \\ \hline
Temporal attributes  & 53 (41) & 76 (49) & \textbf{84 (66)} & 80 (57) & 70 (51) & 64 (29) \\ \hline
Frozen attributes  & 56 (56) & 68 (68) & \textbf{82 (82)} & 75 (75) & 68 (68)& 57 (57) \\ \hline
Dynamic evolution   & \textbf{79 (74)} & \textit{63(56)} & \textit{68 (55)} & 66 (56) & 70 (62) & 71 (57)\\ \hline
Dynamic extension    & \textbf{86 (58)} & 66 (54) &63 (31)  & 65 (43) & 72 (48) & n/a \\ \hline
Transition in the past         & 65 (60) & 56 (54) &53 (45) & 57 (51) & 60 (54) & 29 (21) \\ \hline
 Both optional and mandatory transitions         & 65 (65)  & 56 (56) & 61 (53)  & 58(54) & 61 (59) & n/a \\ \hline
\end{tabular}	
\end{table}

\subsubsection{Can novices create \ourERT diagrams?}
E11 specifically had a controlled setting for what needed to be modelled, eliminating the possibility that the participants knew too little of the universe of discourse. This was achieved by creating the specification from 
the controlled natural language sentences generated from the model answer. Groups 1-42 of UG3 were given the {\sc Dev/Dex} variant and groups 43-84 were given the {\sc Chg/Ext} variant. In both sets, one group did not submit, hence, we have 41 models in M-UG3-D and 42 in M-UG3-C. The average total marks were 4.35 and 4.5 out of 6 (or 72.6 and 75 out of a scale of 100), respectively; see Table~\ref{fig:Setcdcnl}. Both sets had a $>95\%$ correct on the clock/no clock (column B in Table~\ref{fig:Setcdcnl}), and to have at least one of the optional transitions (column G), and for M-UG3-C, the clock on the attribute Company as well (column F). Only about half had all three evolution constraints correct (column K),  with M-UG3-C better than M-UG3-D (59.52\% vs 51.22\%), and less than half had all transitions in the past correct (column I), again with M-UG3-C better (40.48\% in M-UG3-C vs. 34.15\% for M-UG3-D). The only time that M-UG3-C scored substantially lower was for the clocks on the relationships (66.67\% vs 73.17\%, column D). Overall, M-UG3-C scored better than M-UG3-D.

Since we are interested in the transitions, we checked whether the correct modelling of the transitions depends on how well they get the easier non-transition temporal aspects correct. We aggregated the latter (B-F in Table~\ref{fig:Setcdcnl}) and the former (G-M in Table~\ref{fig:Setcdcnl}) and carried out a paired t-test for both sets (after calibrating to the same scale, by dividing by 5, resp. 7): for M-UG3-D, p=0.00026, and for M-UG3-C, p=0.00382, i.e., the difference is statistically significant. That is, the scores for the temporal constraints on the elements are statistically significantly higher than the same model's transition constraints, where more (fewer) errors with the easier elements also means more (resp. fewer) errors in the transition constraints, and that for M-UG3-D slightly more so (with 0.82 vs 0.66 out of 1, i.e. 0.16 less) than for M-UG3-C (with 0.85 vs 0.72 out of 1, i.e., 0.13 less).

\begin{table}[t]
\small
	\centering
	\caption{Scores in percentages (rounded) over the two sets of models (for B-M) and the average for each set (N), in E11.
B:	Traveller, PrevCust and VIPCust have a clock;
C:	Client and Flight do NOT have a clock;
D:	a clock on any of those 3 relationships;
E:	Arrival has pin, and nothing else in the entire model has a pin;
F:	company has clock and delay has no clock;
G:	any one of those 4 dotted arrows is dotted;
H:	any 1 ``past" transition has a minus; 
I:	all 3 above correct; 
J:	any one {\sc Dev} has that word (irrespective of minus/not); 
K:	all 3 above correct; 
L:	any one {\sc Dex} has that word (irrespective of minus/not); 
M:	all 6 above correct  (i.e.have that word, irrespective of whether minus is there or not).
	} \label{fig:Setcdcnl}
		\begin{tabular}{|p{1.4cm}|p{0.45cm}|p{0.45cm}|p{0.45cm}|p{0.45cm}|p{0.45cm}|p{0.45cm}|p{0.45cm}|p{0.45cm}|p{0.45cm}|p{0.45cm}|p{0.45cm}|p{0.45cm}|p{0.45cm}|}
\hline
\textbf{ } & \textbf{B} & \textbf{C} & \textbf{D} &  \textbf{E} &  \textbf{F} &  \textbf{G} &  \textbf{H}&  \textbf{I}&  \textbf{J}&  \textbf{K}&  \textbf{L}&  \textbf{M}&  \textbf{N}  \\ \hline \hline
M-UG3-D & 95 &100 &\textbf{73}& 59 & 85& 100 &81 &34 & 85 & 51 &  83 & 24 & 67  \\ \hline
M-UG3-C & 98 & 95& 67& \textbf{69} & \textbf{95} &100 &79 &\textbf{41} & \textbf{93}& \textbf{60} &\textbf{93} &\textbf{43} & \textbf{83} \\ \hline
\end{tabular}
\end{table}

\section{Discussion}
\label{sec:disc}

The results showed slightly better outcomes for {\sc Chg/Ext} than {\sc Dev/Dex}, especially in the more structured modelling task in E11 (results in Table~\ref{fig:Setcdcnl}), although not convincingly so in E9\footnote{
Specifically regarding the results of E9: it leans toward better results with {\sc Dev/Dex}  than {\sc Chg/Ext} in Table~\ref{fig:useOK}, due to the generally higher accuracy, but with mixed results in self-reported ease/difficulty in Table~\ref{fig:migrateOK}, and slightly clearer {\sc Dev$^-$/Dex$^-$} over lower-case (Fig.~\ref{fig:pastOK}). In contrast,  {\sc Chg/Ext} constraints are used more per model than  {\sc Dev/Dex} (Table~\ref{fig:useOK}), thus contradicting perception, and those models have a higher usage of other temporal features compared to the models that use {\sc Dev/Dex}, notably regarding temporal entity types, more use of temporal relationships, and temporal and frozen attributes.}. 
In both cases, we had used the same algorithm for creating groups, which thus excludes 
the possibility of groups of varying aptitude. 
An average score of in the low 70s (out of 100) based off highly structured sentences is not an excellent mark for modelling with \ourERT. However, t 
they did not do better in the the atemporal, regular, EER diagram creation task of the assignment. Furthermore, 
 it was the last question of the assignment and for relatively few marks. Also, these pass marks were achieved without any teaching or support on temporal conceptual data modelling. The only aide was the few pages of written explanation as part of the assignment. This also suggests that modellers in industry may be able to commence with temporal conceptual modelling without the 
preparatory training
that was needed to create \ourERT in the first place. Put differently: the lessons learned from the reflective teachings approach in experiments E1-E5 have made it into \ourERT to a sufficient extent.

The amount of explanation (M-UG1 versus M-UG2) did not have the expected effect of better results both in perception of understanding and quality of the models developed. While this may be due to saturation of the explanation, the extended brief does have additional explanation on visualising key concepts of the flow of time, a slightly longer description of the \ourERT elements, and a short paragraph illustrating the controlled natural language sentences. Possible interfering factors may have been the different lecturer for UG2 that that may have affected general understanding of ERD (1st time junior lecturer to UG2 compared to seasoned professor to UG1 who had taught ERD many times) and  E8 was conducted during COVID-19's `physically distanced learning' in 2021 (i.e., online teaching \& learning) and the campus fire around the same time had disturbed the students. No substantive decline in accuracy between the models in M-UG1 versus M-UG2 may thus indicate an improved effect thanks to the extended explanation.

Threats to validity of the results are negligible for the large-scale experiment thanks to the size of both the number of students in each of E8, E9 and E11 and that varying parameters and circumstances yielded comparable results. A possible threat to validity may be larger for E1-E5, since they were largely qualitative with few participants. On the positive side, however, they were in-depth and gave insight into the level of difficulty of mastering TCDM that one could not have obtained with quantitative studies alone. Such a series of experiments with other business users might result in a different narrowing down of icon preferences. However, the scaling-up of asking feedback about notation in E6-E8 and their preferences do strongly suggest the chosen notation is acceptable.

The only other experiment carried out with TCDMLs are those we reported in \cite{KB17}, where we compared it to set-based semantics notation, Description Logics notation in $\dlrus$, a coding-style notation, controlled natural language sentences, and hand-drawn sketches of \ourERT (with {\sc Dev/Dex}). The clear preference of the graphical notation there held also for the much larger group of participants in E8 and E9 (Fig.~\ref{fig:langPrefs}).

Another threat to the validity might be that there was no existing modelling tool for \ourERT, requiring the participants to add the elements manually themselves rather than through drag-'n-drop features. We observed true syntactic mistakes only rarely, such as an \EXT\, from the entity type {\sf Course} to an attribute {\sf requirements} of an entity type {\sf Major}, and ample use of temporal constraints, and therefore it is unlikely to affect the overall conclusions.

Returning to our research questions from the Introduction section, we can answer them, as follows. For RQ1, on what diagram notation is preferred for temporal elements and constraints, this culminated in the final \ourERT notation. The evaluations provided evidence for RQ2 and RQ3,  on how well the resulting temporal conceptual model is understood by modellers and how well   they are able to design such models, respectively. The answers to these are, overall, `well enough', hovering around 67-98\% correctness in the controlled domain (Table~\ref{fig:Setcdcnl}). The amount of mistakes made are not worse than in atemporal EER diagrams.

Lastly, now that there is an evidence-based very expressive TCDML that is usable by novice modellers, it may merit the effort to implement it in a modelling tool and from there to devise algorithms to convert the temporal constraints of \ourERT into constraints in the physical schema of the database. Another possible direction could be to create fragments of \ourERT to match the DLs of temporal ontology-based data access \cite{Artale14,Kalayci19}. 
Such a suitable decidable fragment then could also serve to add a feature to computationally verify the semantic coherence and correctness of the temporal and atemporal constraints and deduce implicit constraints. The demonstrated usability of \ourERT may provide sufficient demand to pursue that line of research.

\section{Conclusions and Future Work}
\label{sec:concl}

The paper presented the new, to the best of our knowledge most expressive, temporal conceptual data modelling language \ourERT. It emerged from a rigorous process of language development through qualitative focus-groups to identify key icons for the temporal constructs and six subsequent quantitative experiments that finalised and solidified the notation and assessed understandability and usability by modellers. The large-scale experiments demonstrated that the labels for the transition constraints do not  
affect model correctness much, and providing an extended written explanation of \ourERT
also did not improve quality. 
Better quality models were produced when the domain was more precisely described, which emphasises the need for 
guidance for domain experts to describe their needs. 
While extensive training was needed initially, the large scale experiments showed that the resultant \ourERT does 
result 
in models of sufficient quality 
from 
novice modellers, albeit leaving room for improvement.

Many future research directions are possible, from tooling to spin-offs for temporal ontology-based data access to reaching out to easing the development of temporal ontologies.

\section*{Author Contributions}
TS carried out the literature review and Experiments E1-E5, supervised by CMK. SB and CMK designed and executed E6-E11, data analysis of E7, E8, E9, E11 was carried out by SB, E10's analysis by CMK, and E6-E11 was discussed jointly by SB and CMK. The formal definition of \ourERT was specified by CMK. CMK coordinated the writing of the manuscript.  

\section*{Acknowledgements}
We would like to thank all participants in the experiments.

\bibliographystyle{elsarticle-num}
\bibliography{trendRefs}

\section*{Appendix}

\subsection*{Appendix A1 -- Examples of alternative notations for data involving time}

\newpage
\subsection*{Appendix A2 -- Diagrams used in the notation experiments}

\begin{figure}[h]
\centering
\includegraphics[width = 0.9\textwidth]{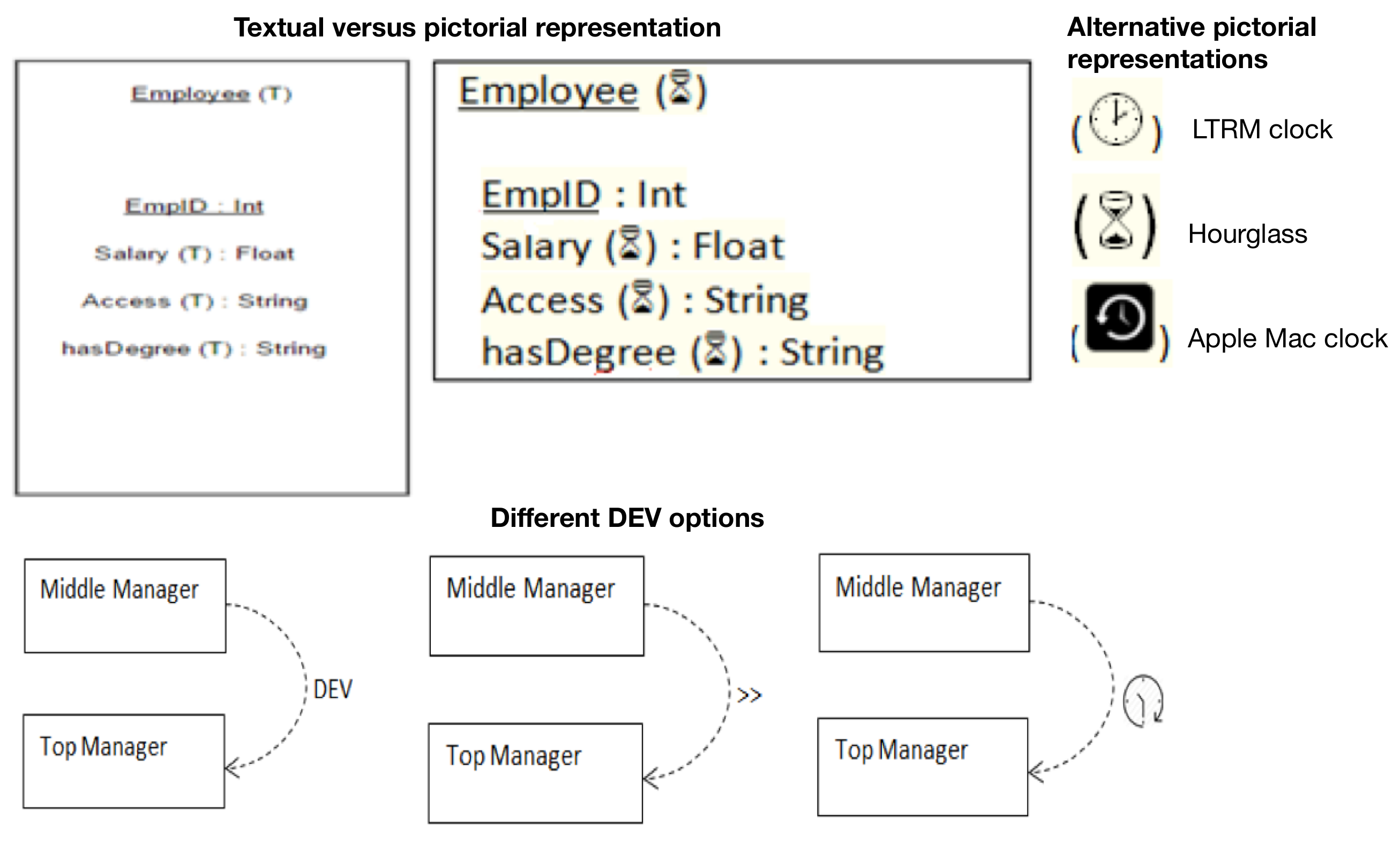}
\caption{Representative selection of questions about alternative representations of temporal information. For the textual representation, it also included a question on preference for a specific colour, and for {\sc Dev}, the other options were ``$>=$'' and two arrows making a circle-shape with a dot in the middle.}
\label{fig:Exp1qs}
\end{figure}

\begin{figure}[h]
\centering
\includegraphics[width = 0.9\textwidth]{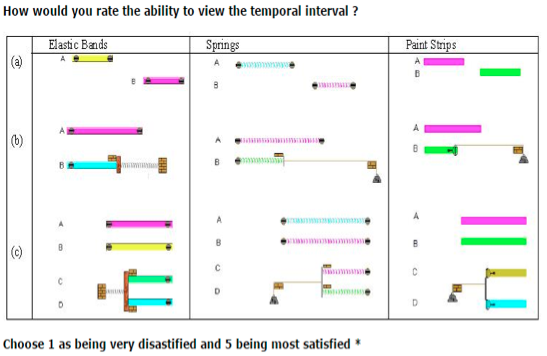}
\caption{One of the questions about springs, elastic bands, and paint strips; the same figure was used for ``which proposal shows least clarity in terms of the changing period?'' where participants had to choose out of spring, elastic band, and paint strip.}
\label{fig:Exp2q}
\end{figure}

\begin{figure}[h]
\centering
\includegraphics[width = 0.8\textwidth]{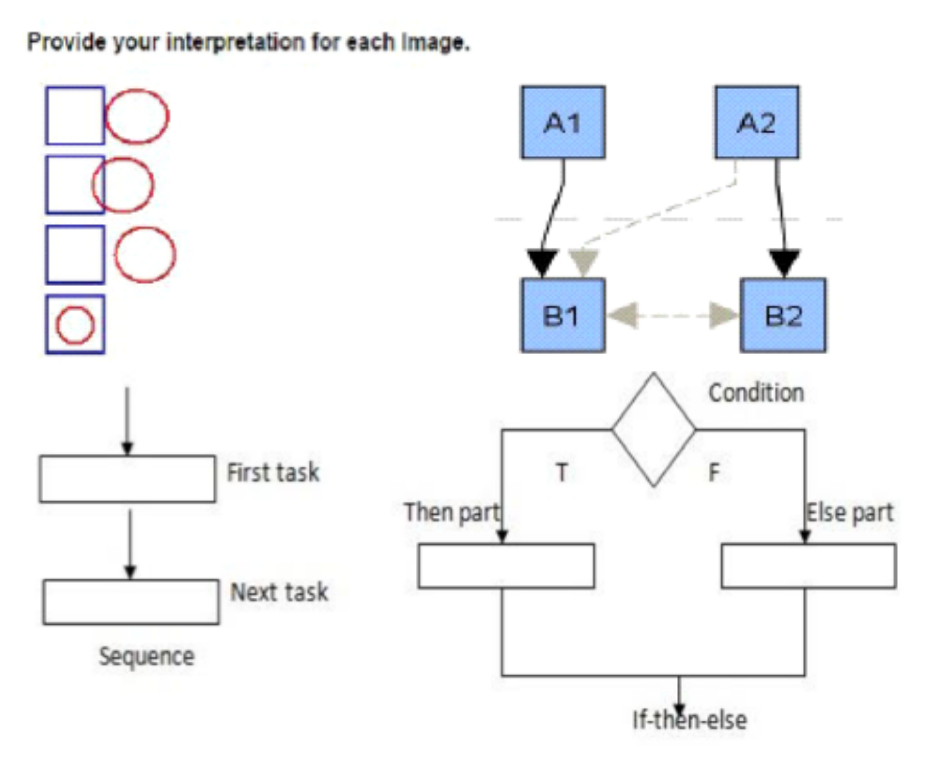}
\caption{Diagrams of Experiment 3 that participants had to interpret.}
\label{fig:Exp3q}
\end{figure}

\begin{figure}[h]
\centering
\includegraphics[width = 0.9\textwidth]{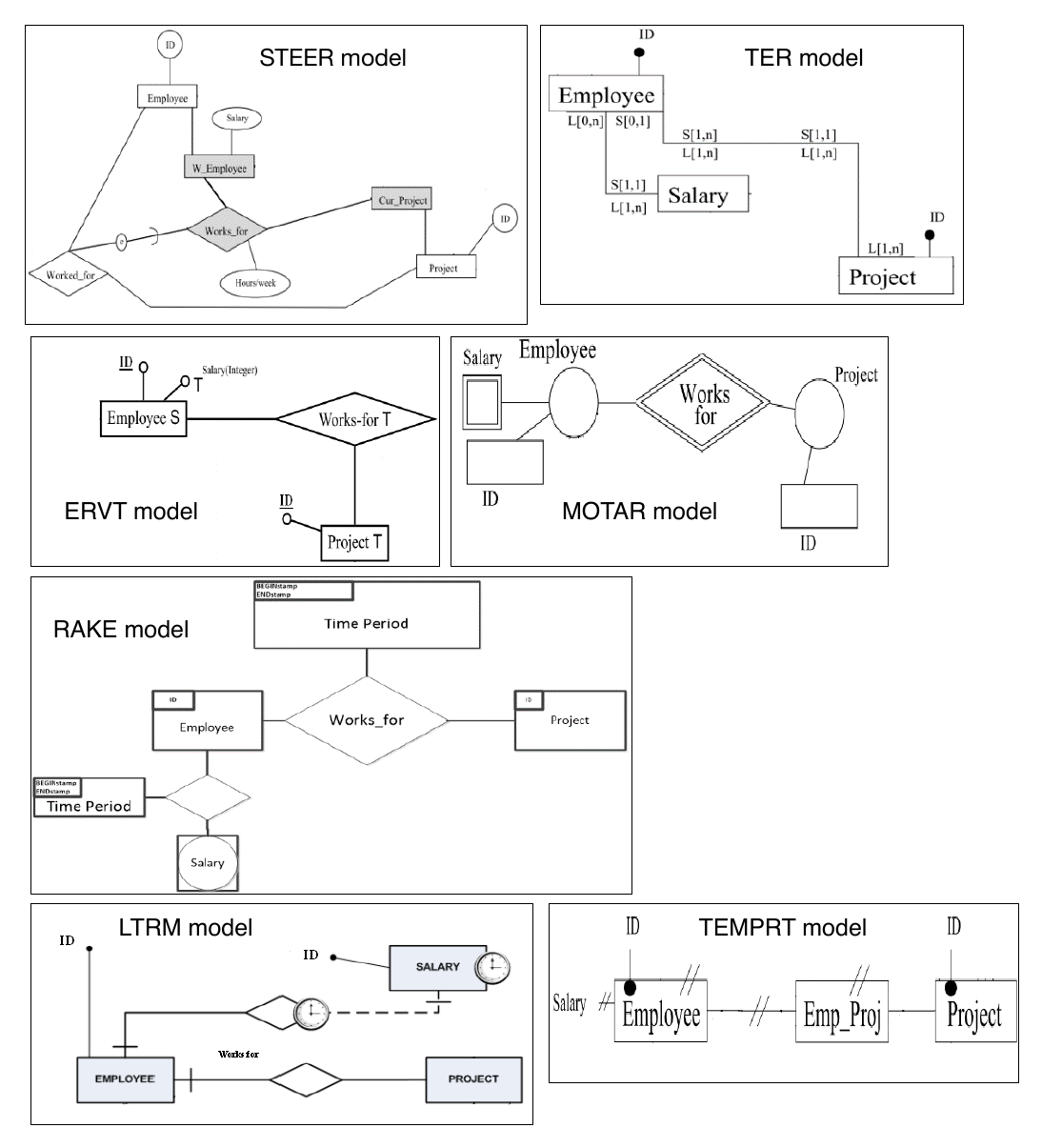}
\caption{Representation of the same temporal information in six different temporal conceptual modelling languages, which participants had to rate.}
\label{fig:Exp5q1}
\end{figure}

\clearpage
\newpage

\subsection*{Appendix A3 -- Summary table for E7-E11}

\vspace{1cm}

\subsection*{Appendix A4 -- Story text and model answer for E11}
\label{sec:e11in}

\noindent \textbf{Draw a TREND data model for the following description:}\\
Flight is an entity type whose objects will always be a flight.\\
Client is an entity type whose objects will always be a client.\\
Traveller is a client.\\
Previous-customer is a client.\\
VIP-customer is a client.\\
Each traveller is not a traveller for some time.\\
Each previous-customer is not a previous-customer for some time.\\
Each VIP-customer is not a VIP-customer for some time.\\
A client may also become a traveller.\\
A client may also become a VIP-customer.\\
Each traveller must evolve to a previous-customer ceasing to be a traveller.\\
A previous-customer may also become a traveller.\\
A previous-customer may also become a VIP-customer.\\
Each VIP-customer was already a previous-customer.\\
Each VIP-customer was already a traveller.\\
A traveller books a flight\\
A traveller pays for a flight.\\
A previous-customer took a flight.\\
Each traveller books a flight will be followed by traveller pays for flight after exactly 3 days, terminating the
traveller books a flight relation.\\
Each previous-customer took a flight must have been preceded by traveller pays for flight, and terminating
that traveller pays for flight relation.\\
Each object in entity type client having attribute company does not have a company at some time.\\
Each object in entity type flight having attribute delay has delay at all times.\\
Once the value for arrival time is set, it cannot change anymore.\\

\begin{figure}[h]
\centering
\includegraphics[width = 0.7\textwidth]{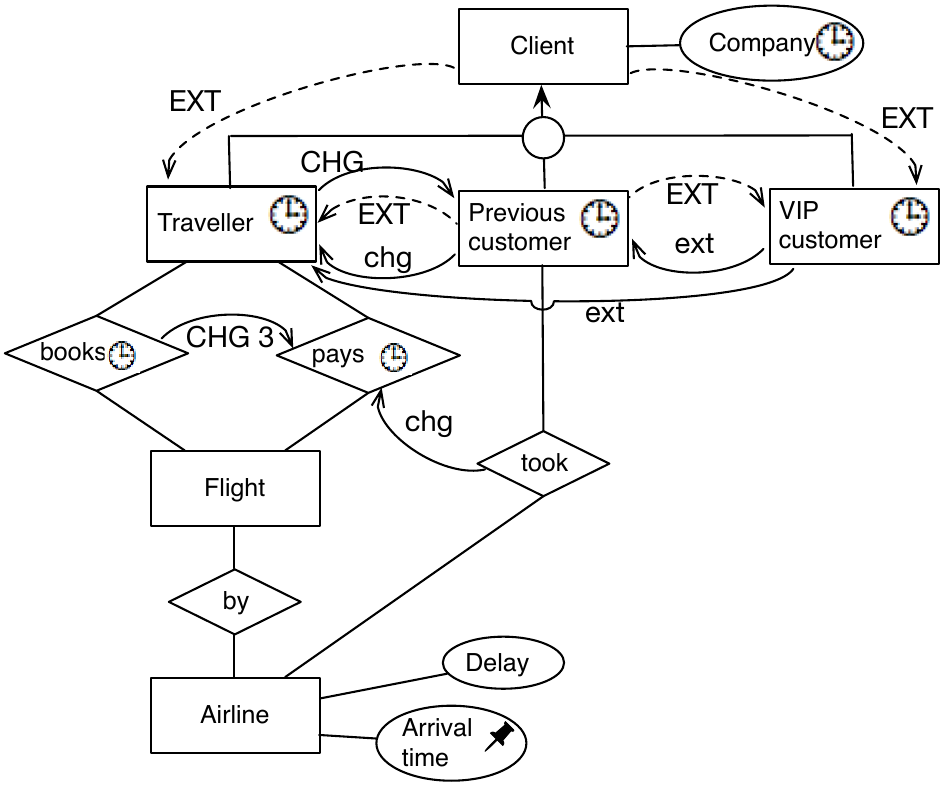}
\caption{Model solution for Experiment 11.}
\label{fig:e11model}
\end{figure}

\vspace{1cm}

\subsection*{Appendix A5 -- Syntax and semantics of \DLRUS}
\label{sec:dlrus}

\begin{figure}[t]
{\tt
{\scriptsize
\begin{center}
\renewcommand{\arraystretch}{1.0}
$\begin{array}{ll}

        C  \to & \top\ \mid\ \bot\ \mid\ C\!N\ \mid\ \neg C\ \mid  C_{1}\ \sqcap\
     C_{2}\ \mid\  \EXISTR{k}{U_j} R\ \mid\ \exists[F] A \mid\\\ &
     \sometimep C\ \mid\ \sometimem C\ \mid\ \alltimep C\ \mid\ \alltimem C\
     \mid\  \nexttime C\ \mid\ \prevtime C\ \mid\ 
     C_1\ \U C_2\ \mid\ C_1\ \S C_2\ \vspace{1.2ex}\\
     R  \to\ & \top_{n}\ \mid\  R\!N
     \ \mid\ \neg R\ \mid R_{1}\ \sqcap \ R_{2}\ \mid\ 
     \selects {U_i/n} C\ \mid\\ &
     \sometimep R\ \mid\ \sometimem R\ \mid\ \alltimep R\ \mid\ \alltimem R\
     \mid\ \nexttime R\ \mid\ \prevtime R\ \mid\ 
      R_1\ \U R_2\ \mid R_1\ \S R_2\ \vspace{1ex} \\

A  \to\ & \top_{A} \mid\  A\!N  \mid\  \neg A \mid\ \selects {F} C  \mid\ \\&
\sometimep A\ \mid\ \sometimem A\ \mid\ \alltimep A\ \mid\ \alltimem A\ 
\mid\ \nexttime A \mid\ \prevtime A \mid A_1 \U A_2 \mid\  A_1\S A_2
\end{array}$
\renewcommand{\arraystretch}{1}
\end{center}
\vspace{-5ex}
}}

{\scriptsize
\renewcommand{\arraystretch}{0.8}

\begin{center}\tt
  {\scriptsize $\begin{array}{r@{\hspace{1ex}}l@{\hspace{.3ex}}}
      \Intt{\top} = & \acto\\
      \Intt{\bot} = & \emptyset\\
      \Intt {C\!N} \subseteq & \Intt{\top}\\
      \Intt{(\neg C)} = & \Intt{\top} \setminus \Intt C\\
      \Intt{(C_{1}\sqcap C_{2})} = & \Intt C_{1} \cap \Intt C_{2}\\
      \Intt{(\EXISTR{k}{U_j} R)} = & \{\ \parbox[t]{\textwidth}{$
        o\in\Intt{\top}\mid \sharp\set{\langle o_{1},\dots,
          o_{n}\rangle\in\Intt R\mid o_{j}=o} \lessgtr k \}
        $}\\
          \Intt{(\exists\,[F] A} = & \{\ \parbox[t]{\textwidth}{$
        o\in\Intt{\top}\mid \sharp\set{\langle o,d\rangle \in\Intt A \geq 1  \} 
       }$}\\
      \Intt{(C_{1}\U C_{2})} = & \{\ \parbox[t]{\textwidth}{$
        o\in\Intt{\top}\mid \exists v>t\per (o\in \Intv{C_{2}}\land
        \forall w\in (t,v)\per o\in
        \Intw{C_{1}})\}$} \\
      \Intt{(C_{1}\S C_{2})} = & \{\ \parbox[t]{\textwidth}{$
        o\in\Intt{\top}\mid \exists v<t\per (o\in \Intv{C_{2}}\land
        \forall w\in (v,t)\per o\in \Intw{C_{1}})\}$}
\vspace{1ex}\\
   \Intt{(\top_{n})} = & (\acto)^{n}\\
   
   \Intt {R\!N} \subseteq & \Intt{(\top_{n})}\\
   \Intt{(\neg R)} = & \Intt{(\top_{n})} \setminus \Intt R\\
   \Intt{(R_{1}\sqcap R_{2})} = & \Intt R_{1} \cap \Intt R_{2}\\
 \Intt{(\selects {U_i/n} C)} = &
       \{\ \parbox[t]{\textwidth}{$
       \langle o_{1},\dots, o_{n}\rangle\in\Intt{(\top_{n})}
       \mid  o_i\in\Intt C\} $} \\

   \Intt{(R_{1}\U R_{2})} = &
       \{\ \parbox[t]{\textwidth}{$
        \langle o_{1},\dots, o_{n}\rangle\in\Intt{(\top_{n})}
        \mid  \exists v>t\per (\auf o_{1},\dots, o_{n}\zu \in
        \Intv{R_{2}}\land \\
                     \forall w\in (t,v)\per \auf o_{1},\dots, o_{n}\zu
       \in \Intw{R_{1}})\}$} \\
\Intt{(R_{1}\S R_{2})} = &
       \{\ \parbox[t]{\textwidth}{$
        \langle o_{1},\dots, o_{n}\rangle\in\Intt{(\top_{n})}
        \mid  \exists v<t\per (\auf o_{1},\dots, o_{n}\zu \in
        \Intv{R_{2}}\land \\
        		\forall w\in (v,t)\per \auf o_{1},\dots, o_{n}\zu
       \in \Intw{R_{1}})\}$} \\
       
\Intt{(\sometimep R)} = & \{\auf o_1,\dots, o_n\zu\in \Intt{(\top_n)}\mid
\exists v>t\per \auf o_1,\dots, o_n\zu\in \Intv{R}   \} \\
\Intt{(\nexttime R)} = & \{\auf o_1,\dots, o_n\zu\in \Intt{(\top_n)}\mid
\auf o_1,\dots, o_n\zu\in R^{\mathcal{I}(t+1)}   \} \\
\Intt{(\sometimem R)} = & \{\auf o_1,\dots, o_n\zu\in \Intt{(\top_n)}\mid
\exists v<t\per \auf o_1,\dots, o_n\zu\in \Intv{R}   \} \\
\Intt{(\prevtime R)} = & \{\auf o_1,\dots, o_n\zu\in \Intt{(\top_n)}\mid
\auf o_1,\dots, o_n\zu\in R^{\mathcal{I}(t-1)} \vspace{1.5ex}\\
   \Intt{(\top_{A})} = & \acto \times \act\\
      \Intt {A\!N} \subseteq & \Intt{(\top_A)}\\
 \Intt{(\selects {F} C)} = &
     \{\ \parbox[t]{\textwidth}{$
       \langle o,d\rangle\in\Intt{(\top_A)}
       \mid  o\in\Intt C\} $} \\
       
   \Intt{(A_{1}\U A_{2})} = &
       \{\ \parbox[t]{\textwidth}{$
        \langle o,d\rangle\in\Intt{(\top_A)}
        \mid  \exists v>t\per (\langle o,d\rangle \in
        \Intv{A_{2}\land }
         \forall w\in (t,v)\per \langle o,d\rangle
       \in \Intw{A_{1}})\}$} \\
\Intt{(A_{1}\S A_{2})} = &
       \{\ \parbox[t]{\textwidth}{$
        \langle o,d\rangle\in\Intt{(\top_A)}
        \mid  \exists v<t\per (\langle o,d\rangle \in
        \Intv{A_{2}}\land  \forall w\in (v,t)\per \langle o,d\rangle
       \in \Intw{A_{1}})\}$} \\
\Intt{(\sometimep A)} = & \{\langle o,d\rangle\in \Intt{(\top_A)}\mid
\exists v>t\per \langle o,d\rangle\in \Intv{A}   \} \\
\Intt{(\nexttime A)} = & \{\langle o,d\rangle\in \Intt{(\top_A)}\mid
\langle o,d\rangle\in A^{\mathcal{I}(t+1)}   \}  \\
\Intt{(\sometimem A)} = & \{\langle o,d\rangle\in \Intt{(\top_A)}\mid
\exists v<t\per \langle o,d\rangle\in \Intv{A}  \} \\
\Intt{(\prevtime A)} = & \{\langle o,d\rangle\in \Intt{(\top_A)}\mid
\langle o,d\rangle\in A^{\mathcal{I}(t-1)}  \}
\end{array}
$}
\end{center}
\vspace{-3.5ex}
}

\caption {Syntax and semantics of $\DLRUS$; $o$ denote objects, $d$ domain values, $v, w, t \in \TSS$.} 
\label{fig:dlrus_semantics}

\end{figure}

This appendix provides an overview of \DLRUS \cite{Artale02}, it being the foundation for both the semantics and DL notation.  \DLRUS  is a fragment of first order logic that combines the propositional temporal logic with \emph{Since} and \emph{Until} operators with the (atemporal) DL $\mathcal{DLR}$~\cite{Calvanese03a} in such a way that they can be used with relationships, entity types, and attributes; its syntax and semantics are included in Fig.~\ref{fig:dlrus_semantics}. As usual for DLs, there are concepts $C$ (atomic ones denoted with $CN$), $n$-ary DL roles $R$ (relationships, with $n \geq 2$, $RN$), attributes $A$ between a class and a datatype, and DL role components ($U$, of which $F$ denotes a role component in an attribute, $F \subseteq U,$ and $ F = \{{\tt From, To}\} $). 
The selection expression $U_i/n : C$ denotes an $n$-ary relation whose $i$-th argument ($i \leq n$) is of type $C$ 
 and $[U_j]R$ denotes the $j$-th argument ($j \leq n$)---i.e., a DL role component, alike a projection over the role---in role $R$ (we omit subscripts $i$ and $j$ if it is clear from the context). 
$\U$ntil and $\S$ince together with $\bot$ and $\top$ suffice to define the relevant temporal operators: $\Diamond^+$ (`some time in the
future') as $\sometimep C \equiv \top\U C$, $\nexttime$ (`at the next moment') as $\nexttime C \equiv \bot\U C$, and likewise for their past counterparts. Analogously, there are  
$\Box^+$ (`always in the future') and $\Box^-$ (`always in the past') as the duals of $\Diamond^+$ and $\Diamond^-$.  
The operators $\sometimes $(`at some moment') and its dual $\alltimes $(`at all moments') are defined as $\sometimes C \equiv C \sqcup \sometimep C \sqcup \sometimem C $ and $\alltimes C \equiv C \sqcap \alltimep C \sqcap \alltimem C$,
respectively.

$\DLRUS$'s model-theoretic semantics is built on the assumption of a linear flow of time $\TS = \langle \TS_p, <\rangle, $\,where\,$ \TS_p$ is a set of countably infinite time points (called chronons) and $<$ is isomorphic to the usual ordering on the integers. The language of $\DLRUS$ is interpreted in temporal models over \TSS, 
which are triples in the form $\IS = \langle\TSS,\acta,\cdot^{\I(t)}\rangle$, where $\Delta^\I$ is the union of two non empty disjoint sets, the {\em domain of objects}, $\acto$, and {\em domain of values}, $\act$, and $\cdot^{\I(t)}$ the interpretation function such
that, for every $t\in{\TSS}$, 
every class $C$, and every $n$-ary relation $R$, we
have $C^{\mathcal{I}(t)}\subseteq\acto$ and
$R^{\mathcal{I}(t)}\subseteq (\acto)^n$; also, $(u,v)=\{w\in\TSS \mid u<w<v\}$. 
A {\em knowledge base} is a finite set $\Sigma$ of \DLRUS\ axioms of
the form $C_{1}\IsSubs C_{2}$ and $R_{1}\IsSubs R_{2}$, and with
$R_{1}$ and $R_{2}$ being relations of the same arity. An
interpretation $\I$ satisfies $C_1\IsSubs C_2$ ($R_{1}\IsSubs R_{2}$)
if and only if the interpretation of $C_1$ ($R_1$) is included in the
interpretation of $C_2$ ($R_2$) at all time, i.e. $\Intt C_1 \subseteq
\Intt C_2$ ($\Intt R_1 \subseteq \Intt R_2$), for all $t\in\TSS$.

This language can be used to represent temporal entity types, relationships, and attributes, and, importantly, also transition constraints, i.e., constraints on evolving objects, relations, and attributes. For the former, one may use either the notation of the \DLRUS semantics directly, or the DL notation of \DLRUS; e.g., $o \in \Bintt{Plant} \rightarrow \forall t'. o \in \Bintx{C}{t'}$ (with $t,t' \in \TSS$) states that object $o$ is a member of the temporal interpretation (indicated with ``$\mathcal{I}(t)$'') of the concept $Plant$ at time $t$, and if true, then (i.e., ``$\rightarrow$'') for all times $t'$ in the set of time points $\TSS$, $o$ is still a member of $Plant$; that is, $o$ is an instance of plant at all time points in the past, present, and future. In \DLRUS notation, this is declared as ${\sf Plant \sqsubseteq \alltimes Plant}$. In contrast, $o \in \Bintt{Professor} \rightarrow \exists t' \neq t. o \notin \Bintx{Professor}{t'}$ states that there is a time $t'$ (different from time $t$) where an object is not a professor, though at time $t$ it was, is, or will be. In \DLRUS notation, this is ${\sf Student \sqsubseteq \sometimes \neg Professor}$. 

The two key transition constraints are {\em dynamic extension} ({\sc Dex}, or {\sc Chg} in our experiments) and {\em dynamic evolution} ({\sc Dev}, or {\sc Ext} in our experiments). In a dynamic extension, the element (object, relation, attribute) is {\em also} an instance of the other entity type whereas with dynamic evolution, the element {\em ceases} to be an instance of the source element. For instance, there may be an extension ${\sf Student \sqcap \neg TeachingAssisant \sqcap \nexttime TeachingAssisant}$, and an evolution is ${\sf Tadpole \sqcap \neg Frog \sqcap \nexttime (\neg Tadpole \sqcap Frog)}$. It is permissible to use shorthand notation for these constraints, as proposed 
in \cite{Artale07a}: {\sc Dex}$_{{\sf Student,TeachingAssisant}}$ and {\sc Dev}$_{{\sf Tadpole,Frog}}$, respectively.

\subsection*{Appendix A5 -- Logic-based reconstruction of \ourERT into $\dlrus$}
\label{app:logic:ourERT}

Given the set-theoretic semantics for \ourERT (or, for
that matter, if it were to have been linked to UML class diagrams or ORM), relevant automated reasoning concepts such as satisfiability, subsumption, and derivation of new
constraints by means of logical implication can, and have been, defined rigorously \cite{Artale07a}. Since the formalisation does not strictly extend the logic used, these reasoning services also apply to the logic-based reconstruction of \ourERT.
\begin{definition}[Reasoning Services]\label{def:ervt-reasoning}
  Let $\schema$ be a schema, $C\in\CS$ a class, and $R\in\RS$ a
  relationship. The following modelling notions can be defined:
  \begin{enumerate}
  \item $C$ ($R$) is {\em satisfiable} if there exists a legal
    temporal database state \B for $\schema$ such that
    $\Bintt{C}\neq\emptyset$ ($\Bintt{R}\neq\emptyset$), for some
    $t\in\TS$;
  \item $\schema$ is {\em satisfiable} if there exists a legal
    temporal database state \B for $\schema$ (\B is also said a {\em
      model} for $\schema$);
  \item $C_1$ ($R_1$) is {\em subsumed} by $C_2$ ($R_2$) in $\schema$
    if every legal temporal database state for $\schema$ is also a
    legal temporal database state for ${C_1\isa C_2}$ (${R_1\isa
      R_2}$);
  \item A schema $\Sigma'$ is \emph{logically implied} by a schema
    $\schema$ over the same signature if every legal temporal database
    state for $\schema$ is also a legal temporal database state for
    $\schema'$.
  \end{enumerate}
\end{definition}


We briefly summarise how \DLRUS is able to capture temporal schemas
expressed in \ourERT---see also \cite{Berardi05,Artale06,AK08dl} for more
details.

\begin{definition}[Mapping \ourERT into \DLRUS]\label{mapping}
  Let $\Sigma = (
  \LS, \rel,
  \att, \crd_A, \\ \crd_R, \isa_C, \isa_R, \isa_U, \disj_C, 
   \disj_R, \cover, \as,\at, \ident, \mathcal{E})$    
  be a \ourERT conceptual data model. The \DLRUS knowledge base, $\K$, mapping
  $\Sigma$ is as follows.
  \begin{itemize}
  \item For each $A\in\AS$, then, $A\IsSubs {\tt From}\!:\!\top\sqcap {\tt
      To}\!:\!\top \in \K$;
  \item If $C_1\isa_C C_2 \in \schema$, then, $C_1\IsSubs C_2\in\K$;
  \item If $R_1\isa_R R_2 \in \schema$, then, $R_1\IsSubs R_2\in\K$;
  \item If $\rel(R)=\langle U_1\!:\!C_1,\ldots,U_k\!:\!C_k\rangle
    \in\schema$, then $R \IsSubs
    U_1\!:\!C_1\sqcap\ldots\sqcap U_k\!:\!C_k\in\K$;
  \item If $\att(C) = \langle A_1:D_1,\ldots,A_h:D_h\rangle
    \in\schema$, then, $C\IsSubs \exists[{\tt From}]A_1 \sqcap\ldots \sqcap
    \exists[{\tt From}]A_h \sqcap \forall[{\tt From}](A_1\to {\tt
      To}:D_1) \sqcap\ldots\sqcap \forall[{\tt From}](A_h\to {\tt
      To}:D_h)\in \K$;
  \item If $\crd_R(C,R,U) = (m,n)\in \schema$, then, $C\IsSubs
    \exists^{\geq m}[U]R \sqcap \exists^{\leq n}[U]R\in \K$;
  \item If $\crd_A(C,A,F) = (m,n)\in \schema$, then, $C\IsSubs
    \exists^{\geq m}[{\tt From}]A \sqcap \exists^{\leq n}[{\tt From}]A\in \K$;    
  \item If $\{C_1,\ldots,C_n\}\disj_C C\in\schema$, then $\K$ contains:\\
    $C_1\IsSubs C\sqcap\neg C_2\sqcap\ldots\sqcap\neg C_n$;\\
    $C_2\IsSubs C\sqcap\neg C_3\sqcap\ldots\sqcap\neg C_n$;\\
    $\ldots$\\
    $C_n\IsSubs C$;
  \item If $\{R_1,\ldots,R_n\}\disj_R R\in\schema$, then $\K$ contains:\\
    $R_1\IsSubs R\sqcap\neg R_2\sqcap\ldots\sqcap\neg R_n$;\\
    $R_2\IsSubs R\sqcap\neg R_3\sqcap\ldots\sqcap\neg R_n$;\\
    $\ldots$\\
    $R_n\IsSubs R$;
  \item If $\{C_1,\ldots,C_n\}\cover C\in \schema$, then $\K$ contains:\\
    $C_1\IsSubs C$;\\
    $\ldots$\\
    $C_n\IsSubs C$;\\
    $C\IsSubs C_1\sqcup\ldots\sqcup C_n$;
  \item If $\ident(C)=A$, then, $\K$ contains:\\
    $C \IsSubs \exists^{=1}[{\tt From}]{\alltimes A}$;\\
    $\top \IsSubs \exists^{\leq 1}[{\tt To}](A\sqcap [{\tt From}]:C)$;
  \item If $C\in\CSS$, then, $C \IsSubs (\alltimes C) \in\K$ (similar
    for $R\in\RSS$);
  \item If $C\in\CST$, then, $C \IsSubs (\sometimes \neg C) \in\K$ (similar
    for $R\in\RST$);
  \item If $\langle C,A\rangle \in\as$, then, $C\IsSubs \forall[{\tt
      From}](A\to \alltimes A) \in\K$;
  \item If $\langle C,A\rangle \in\at$, then, $C\IsSubs \forall[{\tt
      From}](A\to \sometimes \neg A) \in\K$.
  \item For the set of transition constraints $\ES$:
  \begin{itemize}
  	\item If \EXT$_{C_1,C_2}$, then $\K$ contains:\\
		\EXT$_{C_1,C_2} \sqsubseteq C_1 \sqcap \neg C_2 \sqcap \nexttime C_2$\\
		and similarly for relationships;
  	\item If \CHG$_{C_1,C_2}$, then $\K$ contains:\\
		\CHG$_{C_1,C_2} \sqsubseteq C_1 \sqcap \neg C_2 \sqcap \nexttime (\neg C_1 \sqcap C_2)$\\
		and similarly for relationships;	
  	\item If \MEXT$_{C_1,C_2}$, then $\K$ contains:\\
		$C_1 \sqsubseteq \sometimep$\EXT$_{C_1,C_2}$ \\ 
		and similarly for relationships;
  	\item If \MCHG$_{C_1,C_2}$, then $\K$ contains:\\
		$C_1 \sqsubseteq \sometimep$\CHG$_{C_1,C_2}$ \\ 
		and similarly for relationships;		
  	\item If \ext$_{C_1,C_2}$, then $\K$ contains:\\
		\ext$_{C_1,C_2} \sqsubseteq C_1 \sqcap C_2 \sqcap \prevtime \neg C_2$\\
		and similarly for relationships;
  	\item If \chg$_{C_1,C_2}$, then $\K$ contains:\\
		\chg$_{C_1,C_2} \sqsubseteq \neg C_1 \sqcap C_2 \sqcap \prevtime (C_1 \sqcap \neg C_2)$\\
		and similarly for relationships;	
  	\item If \mext$_{C_1,C_2}$, then $\K$ contains:\\
		$C_2 \sqsubseteq \sometimem$\EXT$_{C_1,C_2}$ \\ 
		and similarly for relationships;
  	\item If \mchg$_{C_1,C_2}$, then $\K$ contains:\\
		$C_2 \sqsubseteq \sometimem$\CHG$_{C_1,C_2}$ \\ 
		and similarly for relationships;
	\item If \QEXT$^n_{C_1,C_2}$, then $\K$ contains:\\
		\QEXT$^n_{C_1,C_2} \sqsubseteq C_1 \sqcap \neg C_2 \sqcap \nexttime^n C_2$, where $\nexttime^n$ abbreviates $n$ times $\nexttime$, \\
		and similarly for relationships;
  	\item If \QCHG$^n_{C_1,C_2}$, then $\K$ contains:\\
		\QCHG$^n_{C_1,C_2} \sqsubseteq C_1 \sqcap \neg C_2 \sqcap \nexttime^n (\neg C_1 \sqcap C_2)$\\ 
		where $\nexttime^n$ abbreviates $n$ times $\nexttime$, \\
 		and similarly for relationships (\QCHGr~etc.) and attributes (\achg);
  	\item and likewise for the past counterparts of the quantitative constraints;
	\item If \PEXT$_{C_1,C_2}$, then $\K$ contains:\\
		$\nexttime\EXT_{C_1,C_2} \sqsubseteq \alltimep C_2$\\
		and similarly for relationships;	
	\item If \PCHG$_{C_1,C_2}$, then $\K$ contains:\\
		$\nexttime\CHG_{C_1,C_2} \sqsubseteq \alltimep C_2$\\
		and similarly for relationships;
  	\item If \frz$_A$, then $\K$ contains:\\
		$C \sqsubseteq \exists [{\tt From}]A \sqcap \alltimep[{\tt From}](A \rightarrow {\tt To}:D) \in \K$			

  \end{itemize}   
    
  \end{itemize}
\end{definition}

\vspace{1cm}

\end{document}